\documentclass[epj,nopacs]{svjour}
\usepackage{graphics,epsfig}

\def\mathbf{\vec}

\def\ca{\c{c}\~{a}}

\def\PSfig#1#2{\centerline{\resizebox{#1\hsize}{!}{\includegraphics*{#2}}}}

\begin{document}

\title{Path Integral Bosonization of the 't Hooft Determinant:
       Quasiclassical Corrections}
\author{Alexander A. Osipov\thanks{On leave from the Joint Institute for
 Nuclear Research, Laboratory of Nuclear Problems, 141980 Dubna,
        Moscow Region, Russia.} and Brigitte Hiller}
\institute{Centro de F\'{\i}sica Te\'{o}rica, Departamento de
         F\'{\i}sica da Universidade de Coimbra, 3004-516 Coimbra, Portugal}
\date{March 3, 2004}

\abstract{The many-fermion Lagrangian which includes the 't Hooft six-quark 
flavor mixing interaction ($N_f=3$) and the $U_L(3)\times U_R(3)$ 
chiral symmetric four-quark Nambu -- Jona-Lasinio (NJL) type interactions 
is bosonized by the path integral method. The method of the steepest 
descents is used to derive the effective quark-mesonic Lagrangian 
with linearized many-fermion vertices. We obtain, additionally to 
the known lowest order stationary phase result of Reinhardt and 
Alkofer, the next to leading order (NLO) contribution arising from 
quantum fluctuations of auxiliary bosonic fields around their 
stationary phase trajectories (the Gaussian integral contribution). 
Using the gap equation we construct the effective potential, from 
which the structure of the vacuum can be settled. For some set of 
parameters the effective potential has several extrema, that in the 
case of $SU(2)_I\times U(1)_Y$ flavor symmetry can be understood on 
topological grounds. With increasing strength of the fluctuations the
spontaneously broken phase gets unstable and the trivial vacuum is 
restored. The effective potential reveals furthermore the 
existence of logarithmic singularities at certain field expectation 
values, signalizing caustic regions.}

\authorrunning{A.A. Osipov and B. Hiller}
\titlerunning{Path integral bosonization of the 't Hooft determinant:
              quasiclassical corrections}
\maketitle

%\pacs{12.39.Fe, 11.30.Rd, 11.30.Qc} 
%%%%%%%%%%%%%%%%%%%%%%%%%%%%%%%%%%%%%%%%%%%%%%%%%%%%%%%%%%%%%%%%%%%%%%%%%%%%%%%
\section{Introduction}
%%%%%%%%%%%%%%%%%%%%%%%%%%%%%%%%%%%%%%%%%%%%%%%%%%%%%%%%%%%%%%%%%%%%%%%%%%%%%%%
The global $U_L(3)\times U_R(3)$ chiral symmetry of the QCD Lagrangian 
(for massless light quarks) is broken by the $U_A(1)$ Adler-Bell-Jackiw 
anomaly of the $SU(3)$ singlet axial current $\bar{q}\gamma_\mu\gamma_5q$.
Through the study of instantons \cite{Hooft:1976,Diakonov:1995},
it has been realized that this anomaly has physical effects with the
result that the theory contains neither a conserved $U(1)$ quantum
number, nor an extra Goldstone boson. Instead, effective $2N_f$ quark
interactions arise, which are known as 't Hooft interactions. In the
case of two flavors they are four-fermion interactions,
and the resulting low-energy theory resembles the original Nambu --
Jona-Lasinio model \cite{Nambu:1961}. In the case of three flavors
they are six-fermion interactions which are responsible for the correct
description of $\eta$ and $\eta'$ physics, and additionally lead to the
OZI-violating effects \cite{Bernard:1988,Kunihiro:1988},
\begin{equation}
\label{Ldet}
  {\cal L}_{2N_f}=\kappa (\mbox{det}\ \bar{q}P_Lq
                         +\mbox{det}\ \bar{q}P_Rq)
\end{equation}
where the matrices $P_{L,R}=(1\mp\gamma_5)/2$ are projectors and the 
determinant is over flavor indices.  

The physical degrees of freedom of QCD at low-energies are mesons. The
bosonization of the effective quark interaction (\ref{Ldet}) by the
path integral approach has been considered in 
\cite{Diakonov:1986,Diakonov:1998}. A similar problem has been 
studied by Reinhardt and Alkofer in \cite{Reinhardt:1988}, where the
$U_L(3)\times U_R(3)$ chiral symmetric four-quark interaction 
\begin{equation}
\label{L4q}
  {\cal L}_4=\frac{G}{2}[(\bar{q}\lambda_aq)^2+
                           (\bar{q}i\gamma_5\lambda_aq)^2]
\end{equation}
has been additionally included to the quark Lagrangian 
\begin{equation}
\label{L_int}
  {\cal L}_{\mbox{int}}={\cal L}_{6}+{\cal L}_4.
\end{equation}
To bosonize the theory in both mentioned cases one has to integrate
out from the path integral a part of the auxiliary degrees of freedom 
which are inserted into the original expression together with 
constraints \cite{Reinhardt:1988} 
\begin{eqnarray}
\label{1}
  1&=&\int \prod_a {\cal D}s_a{\cal D}p_a\delta (s_a-\bar{q}\lambda_aq)
      \delta (p_a-\bar{q}i\gamma_5\lambda_aq)
      \nonumber \\
   &=&\int \prod_a {\cal D}s_a {\cal D}p_a
     {\cal D}\sigma_a {\cal D}\phi_a 
     \\
   &&\exp\left\{i\int d^4x
     [\sigma_a(s_a-\bar{q}\lambda_aq)
    + \phi_a(p_a-\bar{q}i\gamma_5\lambda_aq)]\right\}. \nonumber
\end{eqnarray}
The auxiliary bosonic fields, $\sigma_a$, and, $\phi_a, (a=0,1,\ldots,
8)$ become the composite scalar and pseudoscalar mesons and the auxiliary 
fields, $s_a$, and, $p_a$, must be integrated out. The standard way to do
this is to use the semiclassical or the WKB approximation, i.e. one has to 
expand the $s_a$ and $p_a$ dependent part of the action about the extremal
trajectory.
Both in \cite{Diakonov:1986,Diakonov:1998} and in \cite{Reinhardt:1988} 
the lowest order stationary phase approximation (SPA) has been used to 
estimate the leading contribution from the 't Hooft determinant. In 
this approximation the functional integral is dominated by the
stationary trajectories $r_{\mbox{st}}(x)$, determined by the extremum 
condition $\delta S(r)=0$ of the action $S(r)$\footnote{Here $r$ is a 
general notation for the variables $(s_a, p_a)$ and $S(r)$ is the 
$r$-dependent part of the total action.}. 
The lowest order SPA corresponds to the case in which the integrals 
associated with $\delta^2 S(r)$ for the path $r_{\mbox{st}}(x)$ are 
neglected and only $S(r_{\mbox{st}})$ contributes to the generating 
functional. 
 
In this paper we obtain the $\hbar$-correction to the leading order 
SPA result. It contains not only an extended version of our calculations 
which have been published recently \cite{Osipov:2002} but also includes 
new material with a detailed discussion of analytic solutions of the 
stationary phase equations, calculations of the effective potential to 
one-loop order, solutions of the gap equations, general expressions for
quark mass corrections, and quark condensates. We also discuss the 
results of the perturbative approach to find solutions of the 
stationary phase equations.

There are several reasons for performing the present calculation. First, 
although the formal part of the problem considered here is well known, 
being a standard one-loop approximation, these calculations have never 
been done before. The reason might be the difficulties created by the 
cumbersome structure of expressions due to the chiral group. Special care 
must be taken in the way calculations are performed to preserve the 
symmetry properties of the theory. Second, it provides a nice explicit 
example of how the bosonization program is carried out in the case with 
many-fermion vertices. Third, since the whole calculation can be done 
analytically, the results allow us to examine in detail the chiral 
symmetry breaking effects at the semiclassical level. By including the 
fluctuations around the classical path related with the 't Hooft 
six-quark determinant, our calculations of the gap equations and 
effective potential fill up a gap existing in the literature.
Fourth, the problem considered here is a necessary part of the work 
directed to the systematic study of quantum effects in the extended 
Nambu -- Jona-Lasinio models with the 't Hooft interaction. It has 
been realized recently that quantum corrections induced by mesonic 
fluctuations can be very important for the dynamical chiral symmetry 
breaking \cite{Kleinert:2000,Babaev:2000}, although they are $1/N_c$ 
supressed. 

Let us discuss shortly the main steps of the bosonization which we
are going to do in the following sections. As an example of the 
subsequent formalism, we consider the bosonization procedure 
for the first term of Lagrangian (\ref{Ldet}). Using identity 
(\ref{1}) one has 
\begin{eqnarray}
\label{DP}
 & &\exp\left\{\frac{i}{\hbar}\int d^4x\ \kappa\det (\bar{q}P_Lq)
    \right\} \nonumber \\
   &=&\int\prod_a{\cal D}W_a\exp\left(\frac{i}{\hbar}\int d^4x 
   (-\bar{q}WP_Lq)\right)  \\
   &&\int\prod_a{\cal D}U_a^\dag
   \exp\left\{\frac{i}{\hbar}\int d^4x \left(\frac{1}{2}W_aU^\dag_a
   +\frac{\kappa}{64}\det U^\dag \right)\right\}. \nonumber
\end{eqnarray}
The variables $W=W_a\lambda_a$, where $W_a=\sigma_a-i\phi_a$, 
describe a nonet of meson fields of the bosonized theory. The auxiliary 
variables $U^\dag =U^\dag_a\lambda_a$,
$U^\dag_a=s_a+ip_a$ must be integrated out. The Lagrangian in the
first path integral as well as the Lagrangian $L(W,U^\dag)$ in the
second one have order $N_c^0$, because $\bar{q}q$ and $U^\dag$ count as
$N_c$, $\kappa\sim N_c^{-3}$ and $W\sim N_c^{-1}$. Thus we cannot use 
large $N_c$ arguments to apply the SP method for evaluation of the 
integral. However the SPA is justified in the framework of the
semiclassical approach. In this case the quantum corrections are 
suppressed by corresponding powers of $\hbar$.  
The stationary phase trajectories are given by the equations
\begin{equation}
\label{saddle-point}
  \frac{\partial L}{\partial U^\dag_a}
  =\frac{W_a}{2}+\frac{3\kappa}{64}A_{abc}U^\dag_bU^\dag_c=0
\end{equation}    
where the totally symmetric constants, $A_{abc}$, come from the 
definition of the flavor determinant: 
$$
\det U^\dag=A_{abc}U^\dag_aU^\dag_bU^\dag_c\ ,
$$ 
and equal to
\begin{eqnarray}
\label{A}
   A_{abc}&=&\frac{1}{3!}\epsilon_{ijk}\epsilon_{mnl}(\lambda_a)_{im}
           (\lambda_b)_{jn}(\lambda_c)_{kl} 
           \\
   &=&\frac{2}{3}d_{abc}+\sqrt{\frac{2}{3}}
    \left(3\delta_{a0}\delta_{b0}\delta_{c0}-\delta_{a0}\delta_{bc}
    -\delta_{b0}\delta_{ac}-\delta_{c0}\delta_{ab}\right) \nonumber
\end{eqnarray}
with $\lambda_a$ being the standard $U(3)$ Gell-Mann matrices,
$[\lambda_a,\lambda_b]=2if_{abc}\lambda_c,\
\{\lambda_a,\lambda_b\}=2d_{abc}\lambda_c$,
normalized such that $\mbox{tr}\lambda_a\lambda_b=2\delta_{ab}$, 
and $a=0,1,\ldots ,8$.  

The solution to Eq.(\ref{saddle-point}), $U^\dag_{\mbox{st}}(W)$,
is a function of the $3\times 3$ matrix $W$
\begin{equation}
\label{Ust}
  U^\dag_{\mbox{st}}(W)=4\sqrt{\frac{1}{-\kappa}}W^{-1}(\det 
          W)^{\frac{1}{2}}.
\end{equation} 
Expanding $L(W,U^\dag)$ about the stationary point $U^\dag_{\mbox{st}}(W)$ 
we obtain
\begin{eqnarray}
   L(W,U^\dag )&=&L(W,U^\dag_{\mbox{st}}(W))
               +\frac{1}{2}\tilde{U}^\dag_a
               \frac{\partial^2L}{\partial U^\dag_a\partial U^\dag_b}
               \tilde{U}^\dag_b \nonumber \\
               &+&\frac{1}{3!}\tilde{U}^\dag_a
   \frac{\partial^3L}{\partial U^\dag_a\partial U^\dag_b\partial U^\dag_c}
               \tilde{U}^\dag_b\tilde{U}^\dag_c
\end{eqnarray}
where $\tilde{U}^\dag\equiv U^\dag -U^\dag_{\mbox{st}}$ and, as one
can easily get, 
\begin{eqnarray}
\label{treeres} 
  L(W,U^\dag_{\mbox{st}}(W))&=&\frac{1}{4}\mbox{tr}(W 
         U^\dag_{\mbox{st}})
         +\frac{\kappa}{64}\det U^\dag_{\mbox{st}} \nonumber \\
         &=&2\sqrt{\frac{\det W}{-\kappa}}\ . 
\end{eqnarray}
Therefore, we can present Eq.(\ref{DP}) in the form 
\begin{eqnarray}
\label{DPc}
  && \exp\left\{
     \frac{i}{\hbar }\int d^4x\ \kappa\det (\bar{q}P_Lq)
     \right\}  \\
   &=&\int\prod_a{\cal D}W_a\exp\left\{\frac{i}{\hbar}\int d^4x 
   \left(-\bar{q}WP_Lq+2\sqrt{\frac{\det W}{-\kappa}}\right) 
   \right\} \nonumber \\
 &&\int\prod_a{\cal D}\tilde{U}_a^\dag
   \exp\left\{\frac{i}{\hbar}\int d^4x \left(\frac{1}{2}
   \tilde{U}^\dag_a
   \frac{\partial^2L}{\partial U^\dag_a\partial U^\dag_b}
   \tilde{U}^\dag_b+\ldots
   \right)\right\} \nonumber
\end{eqnarray}
which splits up the object of our studies in two contributions which
we can clearly identify: the first line contains the known tree-level
result \cite{Diakonov:1998} and the second line accounts for the 
$\hbar$-suppressed corrections to it, which we are going to consider. 
Unfortunately, the last functional integral is not well defined. To 
avoid the problem, we will study the theory with Lagrangian 
(\ref{L_int}). In this case the functional integral with quantum 
corrections can be consistently defined in some region ${\cal F}$ 
where the field-independent part, $D_{ab}$, of the matrix 
$\partial^2 L(r)/\partial r_a \partial r_b$ has real and positive 
eigenvalues. In order to estimate the effect of the new contribution 
on the vacuum state we derive the modified gap equation and, 
subsequently, integrate it, to obtain the effective potential $V({\cal
F})$. On the boundary, $\partial {\cal F}$, the matrix $D_{ab}$ 
has one or more zero eigenvalues, $d_{0(k)}(\partial {\cal F})=0$, 
and hence $D_{ab}$ is noninvertible. As a consequence, the effective 
potential blows up on $\partial {\cal F}$. This calls for a more 
thorough study of the effective potential in the neighbourhood of 
$\partial {\cal F}$, since the WKB approximation obviously fails here 
(region of the caustic). Sometimes one can cure this problem going into 
higher orders of the loop expansion \cite{Schulman:1981,Kashiwa:2003}. 
Nevertheless, $V({\cal F})$ can be analytically continued for arguments 
exterior to $\partial {\cal F}$, where $d_{0(k)}$ are negative. In fact, 
because of chiral symmetry, we have two independent matrices 
$D_{ab}^{(1)}$ and $D_{ab}^{(2)}$ associated with the two quadratic
forms $s_aD_{ab}^{(1)}s_b$ and $p_aD_{ab}^{(2)}p_b$ in the exponent of 
the Gaussian integral. The eigenvalues of these matrices are positive
in the regions ${\cal F}_1$ and ${\cal F}_2$ correspondingly, and 
${\cal F}_1\supset {\cal F}_2$. Accordingly, the $SU(3)$ effective 
potential is well defined on the three regions: 
${\cal F}_1=\{d_0^{(1)}, d_0^{(2)}>0\}, \ 
 {\cal F}_2=\{d_0^{(1)}>0, d_0^{(2)}<0\}$, and 
${\cal F}_3=\{d_0^{(1)}, d_0^{(2)}<0\}$  
separated by two boundaries $\partial {\cal F}_2$ and $\partial {\cal F}_1$
where $V\rightarrow +\infty$. It means that the effective potential 
has one stable local minimum in each of these regions. However, we
cannot say at the moment how much this picture might be modified by
going beyond the Gaussian approximation near caustics.
     
Our paper is organized as follows: in Sec. \ref{sec-2} we describe
the bosonization procedure by the path integral for the model with 
Lagrangian (\ref{L_int}) and obtain $\hbar$ corrections to the 
corresponding effective action taking into account the quantum 
effects of auxiliary fields $r_a$. We represent the Lagrangian as a 
series in increasing powers of mesonic fields, $\sigma_a, \phi_a$. 
The coefficients of the series depend on the model parameters $G, 
\kappa , \hat{m}$, and are calculated in the phase where chiral 
symmetry is spontaneously broken and quarks get heavy constituent 
masses $m_u, m_d, m_s$. We show that all coefficients are defined 
recurrently through the first one, $h_a$. 
Close-form expressions for them are obtained in Sec. \ref{sec-3} for 
the equal quark mass as well as $m_u=m_d\neq m_s$ cases. In Sec. 
\ref{sec-3} we also study $\hbar$ corrections to the gap equation. 
We obtain $\hbar$-order contributions to the tree-level constituent 
quark masses. The effective potentials with $SU(3)$ and $SU(2)_I\times
U(1)_Y$ flavor symmetries are explicitly calculated. In Sec. \ref{sec-4} 
we alternatively use the perturbative method ($1/N_c$-expansion) to 
solve the stationary phase equations. We show that this approach leads 
to strong suppression of quantum effects. The result is suppresed by 
two orders of the expansion parameter. We give some concluding remarks 
in Sec. \ref{sec-5}.  Some details of our calculations one can find in 
three Appendices.

%%%%%%%%%%%%%%%%%%%%%%%%%%%%%%%%%%%%%%%%%%%%%%%%%%%%%%%%%%%%%%%%%%%%%%%%%%%%%%%
\section{Path integral bosonization of many-fermion vertices}
\label{sec-2}
%%%%%%%%%%%%%%%%%%%%%%%%%%%%%%%%%%%%%%%%%%%%%%%%%%%%%%%%%%%%%%%%%%%%%%%%%%%%%%%
The many-fermion vertices can be linearized by introducing the
functional unity (\ref{1}) in the path integral representation
for the vacuum persistence amplitude \cite{Reinhardt:1988} 
\begin{equation}
\label{genf1}
   Z=\int {\cal D}q{\cal D}\bar{q}\exp\left(i\int d^4x{\cal L}\right).
\end{equation}  
We consider the theory of quark fields in four dimensional Minkowski 
space, with dynamics described by the Lagrangian density
\begin{equation}
\label{totlag}
  {\cal L}=\bar{q}(i\gamma^\mu\partial_\mu -\hat{m})q
          +{\cal L}_{\mbox{int}}.
\end{equation}
We assume that the quark fields have color $(N_c=3)$ and flavor 
$(N_f=3)$ indices which range over the set $i=1,2,3$. The current 
quark mass, $\hat{m}$, is a diagonal matrix with
elements $\mbox{diag}(\hat{m}_u, \hat{m}_d, \hat{m}_s)$, which 
explicitly breaks the global chiral $SU_L(3)\times SU_R(3)$ symmetry 
of the Lagrangian. The second term in (\ref{totlag}) is given by 
(\ref{L_int}). 

By means of the simple trick (\ref{1}), it is easy to write down the 
amplitude 
(\ref{genf1}) as 
\begin{eqnarray}
\label{genf2}
   Z&=&\int {\cal D}q{\cal D}\bar{q}
       \prod^8_{a=0}{\cal D}s_a
       \prod^8_{a=0}{\cal D}p_a
       \prod^8_{a=0}{\cal D}\sigma_a
       \prod^8_{a=0}{\cal D}\phi_a \nonumber \\
     &&\exp\left(i\int d^4x{\cal L}'\right)
\end{eqnarray}  
with
\begin{eqnarray}
\label{lagr1}
  {\cal L}'&=&\bar{q}(i\gamma^\mu\partial_\mu -\hat{m}-\sigma 
              -i\gamma_5\phi )q
              +\frac{G}{2}\left[(s_a)^2+(p_a)^2\right] \nonumber \\
           &+&s_a\sigma_a+p_a\phi_a
              +\frac{\kappa}{64}\left[
               \mbox{det}(s+ip)
              +\mbox{det}(s-ip)\right]
\end{eqnarray}
where, as everywhere in this paper, we assume that $\sigma
=\sigma_a\lambda_a$, and so on for all auxiliary fields: $\phi ,\ s,\ p$.
Eq.(\ref{genf2}) defines the same expression as Eq.(\ref{genf1}). To see
this, one has to integrate first over auxiliary fields $\sigma_a,\ \phi_a$. 
It leads to $\delta$-functionals which can be integrated out by taking
integrals over $s_a$ and $p_a$ and which bring us back to the  
expression (\ref{genf1}). From the other side, it is easy to rewrite 
Eq.(\ref{genf2}), by changing the order of integrations, in a form 
appropriate to accomplish the bosonization, i.e., to calculate the 
integrals over quark fields and integrate out from $Z$ the unphysical 
part associated with the auxiliary $s_a,\ p_a$ bosonic fields, 
\begin{eqnarray}
\label{genf3}
   Z&=&\int \prod_a{\cal D}\sigma_a{\cal D}\phi_a
          {\cal D}q{\cal D}\bar{q}
          \exp\left(i\int d^4x{\cal L}_q(\bar{q},q,\sigma ,\phi
          )\right) \nonumber \\
     && \int \prod_a{\cal D}s_a{\cal D}p_a
        \exp\left(i\int d^4x{\cal L}_r(\sigma ,\phi ,s,p)\right)
\end{eqnarray}  
where
\begin{eqnarray}
\label{lagr2}
  {\cal L}_q&=&\bar{q}(i\gamma^\mu\partial_\mu -\hat{m}-\sigma 
              -i\gamma_5\phi )q, \\
\label{lagr3}
  {\cal L}_r&=&\frac{G}{2}\left[(s_a)^2+(p_a)^2\right]
              +(s_a\sigma_a+p_a\phi_a) \nonumber \\
            &+&\frac{\kappa}{32}A_{abc}s_a\left(s_bs_c-3p_bp_c
              \right).
\end{eqnarray}
The Fermi fields enter the action bilinearly, we can always integrate
over them, because in this case we deal with a Gaussian integral. 
At this stage one should also shift the scalar fields
$\sigma_a(x)\rightarrow\sigma_a(x)+\Delta_a$ by demanding that the vacuum 
expectation values of the shifted fields vanish $<0|\sigma_a(x)|0>=0$.
In other words, all tadpole graphs in the end should sum to zero, giving 
us the gap equation to fix constants $\Delta_a$. Here $\Delta_a = m_a -
\hat{m}_a$, with $m_a$ denoting the constituent quark masses 
\footnote{The shift by the current quark mass is needed to hit the 
correct vacuum state, see e.g. \cite{Osipov:2001}. The functional 
integration measure in Eq.(\ref{genf3}) does not change under this 
redefinition of the field variable $\sigma_a(x)$.}. 

To evaluate functional integrals over $s_a$ and $p_a$  
\begin{eqnarray}
\label{intJ}
     {\cal Z}[\sigma ,\phi ;\Delta ]&\equiv& 
     {\cal N}\int^{+\infty}_{-\infty}\prod_a{\cal D}s_a{\cal D}p_a
     \nonumber \\ 
     &&\exp\left(i\int d^4x{\cal L}_r(\sigma +\Delta ,\phi ,s,p)\right)
\end{eqnarray} 
where ${\cal N}$ is chosen so that ${\cal Z}[0,0;\Delta ]=1$,
one has to use the method of stationary phase. Following the standard 
procedure of the method we expand Lagrangian ${\cal L}_r(s,p)$ about 
the stationary point of the system $r^a_{\mbox{st}}=(s^a_{\mbox{st}},\ 
p^a_{\mbox{st}})$.  Near this point the Lagrangian ${\cal L}_r(s,p)$
can be approximated by the sum of two terms
\begin{eqnarray}
\label{lagr4}
  {\cal L}_r(\sigma +\Delta , \phi , s, p)
            &\approx& {\cal L}_r(r_{\mbox{st}}) \nonumber \\
            &+&\frac{1}{2}\sum_{\alpha ,\beta }\tilde{r}_\alpha (x)
            {\cal L}''_{\alpha\beta}(r_{\mbox{st}})\tilde{r}_\beta (x)
\end{eqnarray}
where we have only neglected contributions from the third order 
derivatives of ${\cal L}_r(s,p)$. The stationary point,
$r^a_{\mbox{st}}$, is a solution of the equations 
${\cal L}'_r(s,p)=0$ determining a flat spot of the surface ${\cal L}_r(s,p)$:
\begin{equation}
\label{saddle}
  \left\{
         \begin{array}{rcl}
        && Gs_a+(\sigma +\Delta )_a
         +\displaystyle\frac{3\kappa}{32}A_{abc}(s_bs_c-p_bp_c)=0 \\
           && \\
        && Gp_a+\phi_a-\displaystyle\frac{3\kappa}{16}A_{abc}s_bp_c=0.
         \end{array}
  \right.
\end{equation}
This system is well-known from \cite{Reinhardt:1988}. We use 
in Eq.(\ref{lagr4}) symbols $\tilde{r}^\alpha$ for the differences
$(r^\alpha-r^\alpha_{\mbox{st}})$. To deal with the multitude of integrals 
we define a column $\tilde{r}$ with eighteen components 
$\tilde{r}_\alpha =(\tilde{s}_a, \tilde{p}_a)$ and with the real
and symmetric matrix 
${\cal L}''_{\alpha\beta}(r_{\mbox{st}})$ being equal to
\begin{equation}
\label{Qab}
  {\cal L}''_{\alpha\beta}(r_{\mbox{st}})=
  \left(
  \begin{array}{cc}
  G\delta_{ab}+\displaystyle\frac{3\kappa}{16}A_{abc}s_{\mbox{st}}^{c}
  &-\displaystyle\frac{3\kappa}{16}A_{abc}p_{\mbox{st}}^{c}\\
   -\displaystyle\frac{3\kappa}{16}A_{abc}p_{\mbox{st}}^{c}
  &G\delta_{ab}-\displaystyle\frac{3\kappa}{16}A_{abc}s_{\mbox{st}}^{c}
  \end{array}
  \right).
\end{equation} 
The path integral (\ref{intJ}) can now be concisely written as
\begin{eqnarray}
\label{ancJ2}
     {\cal Z}[\sigma ,\phi ;\Delta ]&\approx& {\cal N}\exp\left(i\int d^4x 
     {\cal L}_r(r_{\mbox{st}})           \right)
     \int^{+\infty}_{-\infty}
     \prod_\alpha{\cal D}\tilde{r}_{\alpha} \nonumber \\
     &&\exp\left(\frac{i}{2}\int d^4x\tilde{r}^{\mbox{t}}(x)
     {\cal L}''_r(r_{\mbox{st}})
     \tilde{r}(x)\right).
\end{eqnarray} 
The Gaussian multiple integrals in Eq.(\ref{ancJ2}) define a function 
of ${\cal L}''_{\alpha\beta}(r_{\mbox{st}})$ which can be calculated 
by a generalization of the well-known formula for a one-dimensional 
Gaussian integral. Before we do this, though, some additional comments 
should be made:  

(1) The first exponential factor in Eq.(\ref{ancJ2}) is not new. 
It has been obtained by Reinhardt and Alkofer in \cite{Reinhardt:1988}. A 
bit of manipulation with expressions (\ref{lagr3}) and (\ref{saddle}) leads 
us to the result
\begin{eqnarray}
\label{Lrst}
   {\cal L}_r(r_{\mbox{st}})&=&\frac{1}{6}\left\{
            G[(s_{\mbox{st}}^a)^2+(p_{\mbox{st}}^a)^2]
            +4[(\sigma +\Delta)_as^a_{\mbox{st}}
            +\phi_ap^a_{\mbox{st}}]\right\}
            \nonumber \\
            &=&\frac{G}{12}\mbox{tr}(U_{\mbox{st}}U_{\mbox{st}}^\dag )
              +\frac{1}{6}\mbox{tr}(WU_{\mbox{st}}^\dag
              +W^\dag U_{\mbox{st}}). 
\end{eqnarray} 
Here the trace is taken over flavor indices. We also use the notation 
$W=W_a\lambda_a$ and $U=U_a\lambda_a$ where $W_a=\sigma_a+\Delta_a
-i\phi_a, \ U_a=s_a-ip_a$. It is similar to the notation chosen in 
Eq.(\ref{DP}) with the only difference that the scalar field $\sigma_a$ 
is already splitted as $\sigma_a\rightarrow \sigma_a +\Delta_a$.   
This result is consistent with (\ref{treeres}) in the limit $G=0$.
For this partial case Eq.(\ref{saddle}) coincides with 
Eq.(\ref{saddle-point}) and we know its solution (\ref{Ust}). 
If $G\neq 0$ we have to obtain the stationary point $U_{\mbox{st}}$ from 
Eq.(\ref{saddle}).  
  
One can try to solve Eqs.(\ref{saddle}) exactly, looking for solutions 
$s^a_{\mbox{st}}$ and $p^a_{\mbox{st}}$ in the form of increasing
powers in fields $\sigma_a , \phi_a$  
\begin{eqnarray}
\label{rst}
   s^a_{\mbox{st}}
    &=&h_a+h_{ab}^{(1)}\sigma_b
       +h_{abc}^{(1)}\sigma_b\sigma_c
       +h_{abc}^{(2)}\phi_b\phi_c 
       \nonumber \\
    &+&h_{abcd}^{(1)}\sigma_b\sigma_c\sigma_d
       +h_{abcd}^{(2)}\sigma_b\phi_c\phi_d
       +\ldots \\
   p^a_{\mbox{st}}&=&h_{ab}^{(2)}\phi_b
       +h_{abc}^{(3)}\phi_b\sigma_c
       +h_{abcd}^{(3)}\sigma_b\sigma_c\phi_d
       \nonumber \\
     &+&h_{abcd}^{(4)}\phi_b\phi_c\phi_d
       +\ldots 
\end{eqnarray}
with coefficients depending on $\Delta_a$ and coupling constants.
Putting these expansions in Eqs.(\ref{saddle}) one obtains a series 
of selfconsistent equations to determine $h_a$, $h^{(1)}_{ab}$, 
$h^{(2)}_{ab}$ and so on. The first three of them are
\begin{eqnarray}
\label{ha}
   &&Gh_a+\Delta_a+\frac{3\kappa}{32}A_{abc}h_bh_c=0, \nonumber \\  
   &&\left(G\delta_{ac}+\frac{3\kappa}{16}A_{acb}h_b
     \right)h^{(1)}_{ce}=-\delta_{ae}\ , \\
   &&\left(G\delta_{ac}-\frac{3\kappa}{16}A_{acb}h_b
     \right)h^{(2)}_{ce}=-\delta_{ae}\ . \nonumber
\end{eqnarray}
All the other equations can be written in terms of
the already known coefficients, for instance, we have
$$
   h^{(1)}_{abc}=\frac{3\kappa}{32}h^{(1)}_{a\bar a}h^{(1)}_{b\bar b} 
                 h^{(1)}_{c\bar c}A_{\bar a\bar b\bar c}\ , \quad 
   h^{(2)}_{abc}=-\frac{3\kappa}{32}h^{(1)}_{a\bar a}h^{(2)}_{b\bar b} 
                 h^{(2)}_{c\bar c}A_{\bar a\bar b\bar c}\ , 
$$ 
$$
   h^{(3)}_{abc}=-\frac{3\kappa}{16}h^{(2)}_{a\bar a}h^{(2)}_{b\bar b} 
                 h^{(1)}_{c\bar c}A_{\bar a\bar b\bar c}\ , \quad 
   h_{abcd}^{(1)}=\frac{3\kappa}{16}h^{(1)}_{a\bar a}
                  h^{(1)}_{b\bar b}h^{(1)}_{\bar ccd}
                  A_{\bar a\bar b\bar c}\ , 
$$
\begin{equation}
\label{coeffi}
   h_{abcd}^{(2)}=\frac{3\kappa}{16}h^{(1)}_{a\bar a}
                  \left(h^{(1)}_{b\bar b}h^{(2)}_{\bar ccd}
                  -h^{(2)}_{c\bar b}h^{(3)}_{\bar cdb}\right)
                  A_{\bar a\bar b\bar c}\ , \ \ldots   
\end{equation}
It is assumed that coupling constants $G$ and $\kappa$ are chosen 
such that Eqs.(\ref{ha}) can be solved. 
Let us also give the relations following from (\ref{ha}) which have 
been used to obtain (\ref{coeffi}) 
\begin{equation}
  h_b=(Gh_a+2\Delta_a)h^{(1)}_{ab}
     =-(3Gh_a+2\Delta_a)h^{(2)}_{ab}.
\end{equation}
As a result the effective Lagrangian (\ref{Lrst}) can be expanded in powers 
of meson fields. Such an expansion, up to and including the terms
which are cubic in $\sigma_a, \phi_a$, looks like
\begin{eqnarray}
\label{lam}
   {\cal L}_r(r_{\mbox{st}})
   &=&h_a\sigma_a
      +\frac{1}{2}h_{ab}^{(1)}\sigma_a\sigma_b  
      +\frac{1}{2}h_{ab}^{(2)}\phi_a\phi_b \nonumber \\
   &+&\frac{1}{3}\sigma_a\left[h^{(1)}_{abc}\sigma_b\sigma_c
      +\left(h^{(2)}_{abc}+h^{(3)}_{bca}\right)\phi_b\phi_c\right]
      \nonumber \\
   &+&{\cal O}(\mbox{field}^4).
\end{eqnarray}  
This part of the Lagrangian is responsible for the dynamical symmetry 
breaking in the quark system and for the masses of mesons in the
broken vacuum.

(2) The coefficients $h_a$ are determined by couplings $G, \kappa$
and the mean field $\Delta_a$. This field has in general only three 
non-zero components with indices $a=0,3,8$, according to the symmetry 
breaking pattern. The same is true for $h_a$ because of the first 
equation in (\ref{ha}). It means that there is a system of only three 
equations to determine $h=h_a\lambda_a=\mbox{diag}(h_u,h_d,h_s)$ 
\begin{equation}
\label{saddle-1}
  \left\{
         \begin{array}{rcl}
        && Gh_u+\Delta_u+\displaystyle\frac{\kappa}{16}h_dh_s=0 \\
\\ 
        && Gh_d+\Delta_d+\displaystyle\frac{\kappa}{16}h_sh_u=0 \\
\\        
        && Gh_s+\Delta_s+\displaystyle\frac{\kappa}{16}h_uh_d=0 \\
         \end{array}
  \right. 
\end{equation}
This leads to a fifth order equation for a one-type variable and can be 
solved numerically. For two particular cases, 
$\hat{m}_u=\hat{m}_d=\hat{m}_s$ and $\hat{m}_u=\hat{m}_d\neq\hat{m}_s$, 
Eqs.(\ref{saddle-1}) can be solved analytically, because they are of 
second and third order, correspondingly. We shall discuss this in the 
next section. 

(3) Let us note that Lagrangian (\ref{lagr3}) is a quadratic polynomial
in $p_a(x)$ and cubic with respect to $s_a(x)$. It suggests 
to complete first the Gaussian integration over $p_a$ and only 
then to use the stationary phase method to integrate over $s_a$. In this 
case, however, one breaks chiral symmetry. This is circumvented by
working with the column variable $r_\alpha$, treating the chiral
partners $(s_a, p_a)$ on the same footing. It is easy to check in the
end that the obtained result is in agreement with chiral symmetry. 

Let us turn now to the evaluation of the path integral in Eq.(\ref{ancJ2}).  
After the formal analytic continuation in the time coordinate
$x_0\rightarrow ix_4$, we have\footnote{It differs from the standard 
Wick rotation by a sign. The sign is usually fixed by the
requirement that the resulting Euclidean functional integral is well defined.
Our choice has been made in accordance with the convergence properties
of the path integral (\ref{ancJ2}).} 
\begin{eqnarray}
\label{intK}
     &&K[r_{\mbox{st}}]={\cal N}
     \int^{+\infty}_{-\infty}
     \prod_\alpha{\cal D}\tilde{r}_{\alpha}
     \nonumber \\
     &&\exp\left(-\frac{1}{2}\int d^4x_E\tilde{r}^{\mbox{t}}(x_E)
     {\cal L}''_r(r_{\mbox{st}})
     \tilde{r}(x_E) 
     \right),
\end{eqnarray} 
where the subscripts $E$ denote Euclidean quantities. To find an 
expression for $K[r_{\mbox{st}}]$, we split the matrix 
${\cal L}''_r$ into two parts ${\cal L}''_r(r_{\mbox{st}})=D-L$ 
where
\begin{eqnarray}
\label{L''}
   &&D_{\alpha\beta}=-
             \left(
                   \begin{array}{cc}
                   h^{(1)-1}_{ab}&0\\
                   0&h^{(2)-1}_{ab}
                   \end{array}
      \right)_{\alpha\beta}\ ,\nonumber \\
   &&L_{\alpha\beta}=
             \frac{3\kappa}{16}A_{abc}\left(
                   \begin{array}{cc}
                   -(s^c_{\mbox{st}}-h_c)&
                   p^c_{\mbox{st}}\\
                   p^c_{\mbox{st}}&
                   s^c_{\mbox{st}}-h_c
                   \end{array}
                   \right)_{\alpha\beta}\ .  
\end{eqnarray}
The matrix $D$ corresponds to ${\cal L}''_r$ evaluated at the point 
$s_{\mbox{st}}^a=h_a,\ p_{\mbox{st}}^a=0$ and is simplified with the 
help of Eqs.(\ref{ha}). The field-dependent exponent with matrix $L$ 
can be represented as a series. Therefore, we obtain
\begin{eqnarray}
\label{intK2}
     K[r_{\mbox{st}}]&=&
     {\cal N}\int^{+\infty}_{-\infty}
     \prod_\alpha{\cal D}\tilde{r}_{\alpha}
     \exp\left(-\frac{1}{2}\int d^4x_E\tilde{r}_{\alpha}
     D_{\alpha\beta}\tilde{r}_\beta  
     \right) \nonumber \\
     &&\sum_{n=0}^{\infty}\frac{1}{n!}
     \left(\frac{1}{2}\int d^4x_E\tilde{r}_{\alpha}
     L_{\alpha\beta}\tilde{r}_\beta\right)^n.
\end{eqnarray} 
The real symmetric matrix $D_{\alpha\beta}$ can be diagonalized by 
a similarity transformation $\tilde{r}_\alpha =S_{\alpha\beta}
\tilde{r}'_\beta$. We are then left with the eigenvalues of the matrix 
$D$ in the path integral (\ref{intK2}). These eigenvalues are real and 
positive in a finite region fixed by the coupling constants. For
instance, if $\hat m_u=\hat m_d=\hat m_s$ the region is 
$(-\kappa\Delta_u)<12G^2$ where, as usual in this paper, we assume
that $G>0$ and $\kappa <0$. 
In this region the path integral is a Gaussian one and converges.  
To perform this integration, we first change variables $\tilde{r}'_\beta 
=U_{\beta\sigma}q_\sigma$, such that $U$ rescales eigenvalues
to 1. The quadratic form in the exponent becomes 
$\tilde{r}_\alpha D_{\alpha\beta}\tilde{r}_\beta = 
\tilde{r}'_\alpha (S^tDS)_{\alpha\beta}\tilde{r}'_\beta =q_\alpha 
q_\alpha$. The matrix of the total transformation,
$V_{\alpha\sigma}=S_{\alpha\beta}U_{\beta\sigma}$, has the 
block-diagonal form 
\begin{eqnarray}
\label{matrixV}
    &&V_{\alpha\beta}=
             \left(
                   \begin{array}{cc}
                   V^{(1)}_{ab}&0\\
                   0&V^{(2)}_{ab}
                   \end{array}
      \right)_{\alpha\beta}\ ,\qquad
    h^{(1)}_{ab}=-V^{(1)}_{ac}V^{(1)}_{bc}, \nonumber \\
    &&h^{(2)}_{ab}=-V^{(2)}_{ac}V^{(2)}_{bc}.
\end{eqnarray}
Then the integral (\ref{intK2}) can be written as
\begin{eqnarray}
\label{intK3}
     K[r_{\mbox{st}}]&=&
     {\cal N}\det V\int^{+\infty}_{-\infty}
     \prod_\alpha{\cal D}q_{\alpha}
     \exp\left(-\frac{1}{2}\int d^4x_E q_{\alpha}q_{\alpha}
     \right) \nonumber \\ 
     &&\sum^{\infty}_{n=0}
     \frac{1}{n!}
     \left(\frac{1}{2}\int d^4x_E q_{\alpha}
     (V^tLV)_{\alpha\beta}q_\beta\right)^n.
\end{eqnarray} 

By replacing the continuum of spacetime positions with a discrete
lattice of points surrounded by separate regions of very small
spacetime volume $\Omega$, the path integral (\ref{intK3}) may be 
reexpressed as a Gaussian multiple integral over a finite number of 
real variables $q_{\alpha ,x}$, where $\int d^4x_E\cdots\rightarrow
\Omega\sum_x\cdots$ 
\begin{eqnarray}
\label{intK4}
     K[r_{\mbox{st}}]&=&
     {\cal N}\det V\int^{+\infty}_{-\infty}
     \prod_{\alpha ,x} dq_{\alpha ,x}
     \exp\left(-\frac{\Omega}{2}\sum_{\alpha ,x} q_{\alpha ,x}^2
     \right)
     \nonumber \\
  && \sum^{\infty}_{n=0}
     \frac{1}{n!}
     \left(\frac{\Omega}{2}\sum_{\alpha x, \beta y} q_{\alpha ,x}
     {\cal W}_{\alpha x,\beta y}q_{\beta ,y }\right)^n
\end{eqnarray} 
where the matrix ${\cal W}$ is given by 
${\cal W}_{\alpha x,\beta y}=(V^tLV)_{\alpha\beta}\delta_{x,y}$.
The Gaussian integrals in this expression are well known
\begin{eqnarray}
 &&\int^{+\infty}_{-\infty}
   \prod_{i}^N dq_i \ (q_{k_{1}}q_{k_{2}}\ldots q_{k_{2n}})
   \exp\left(-\frac{\Omega}{2}\sum_i^N q_i^2\right)
  \nonumber \\
 &=&\frac{1}{\Omega^n}\left(\frac{2\pi}{\Omega}\right)^{N/2}
    \delta_{k_{1}k_{2}\ldots k_{2n}}.
\end{eqnarray}  
Here $\delta_{k_{1}k_{2}\ldots k_{2n}}$ is a totally symmetric symbol 
which generalizes an ordinary Kronecker delta symbol, $\delta_{ij}$,
with the recurrent relation
\begin{eqnarray}
  \delta_{k_{1}k_{2}\ldots k_{2n}}&=&
  \delta_{k_{1}k_{2}}\delta_{\hat k_{1}\hat k_{2}\ldots k_{2n}}+
  \delta_{k_{1}k_{3}}\delta_{\hat k_{1}k_{2}\hat k_{3}\ldots k_{2n}}
  \nonumber \\
  &+&
  \delta_{k_{1}k_{2n}}\delta_{\hat k_{1}k_{2}\ldots \hat k_{2n}}.
\end{eqnarray}
The hat in this formula means that the corresponding index must be 
omited in the symbol $\delta$. Let us also remind that integrals of 
this sort with an odd number of $q$-factors in the integrand obviously 
vanish. The multiple index $k$ is understood as a pair $k=\alpha , x$
with the Kronecker $\delta_{k_{1}k_{2}}=\delta_{\alpha_{1}\alpha_{2}}
\delta_{x_{1}x_{2}}$.

By performing the Gaussian integrations one can finally fix 
the constant of proportionality ${\cal N}$ and find that 
\begin{eqnarray}
\label{intK5}
     K[r_{\mbox{st}}]&=&
     1+\frac{1}{2}\delta_{k_{1}k_{2}}{\cal W}_{k_{1}k_{2}}
     +\ldots 
     \nonumber \\
     &+&
     \frac{1}{n!2^n}\delta_{k_{1}k_{2}\ldots k_{2n}}
     {\cal W}_{k_{1}k_{2}}\ldots {\cal W}_{k_{2n-1}k_{2n}}+\ldots 
\end{eqnarray} 
The infinite sum here is nothing else than $1/\sqrt{\mbox{Det}\
(1-{\cal W})}$. The determinant may be reexpressed as a contribution to the 
effective Lagrangian using the relation $\mbox{Det}\ 
(1-{\cal W})=\exp\mbox{Tr}\ln (1-{\cal W})$. 
Thus, we have  
\begin{equation}
\label{intK6}
      K[r_{\mbox{st}}]=
     \exp\left(
     -\frac{1}{2}\mbox{Tr}\ln (1-{\cal W})\right)
\end{equation} 
with the logarithm of a matrix defined by its power series expansion
\begin{eqnarray}
    \ln (1-{\cal W})&=&-{\cal W}-\frac{1}{2}{\cal W}^2
    -\frac{1}{3}{\cal W}^3-\ldots
    \nonumber \\
    &=&
    \delta_{x,y}\left(\ln (1-V^tLV)\right)_{\alpha ,\beta}.
\end{eqnarray}
The path integral (\ref{intK2}) defines a function of $D_{\alpha\beta}$ 
that is analytic in $D_{\alpha\beta}$ in a region around the 
surface where the eigenvalues of $D_{\alpha\beta}$ are real positive 
and the integral converges. Since Eq.(\ref{intK6}) equals to 
(\ref{intK2}), it provides the analytic continuation of Eq.(\ref{intK2}) 
to the whole complex plane, with a cut required by the logarithm. 

Let us now return back from the spacetime discret lattice to the 
spacetime continuum. For that one must take the limit
$N\rightarrow\infty$ in Eq.(\ref{intK6}),
and replace $\sum_x\ldots\rightarrow\Omega^{-1}\int d^4x_E\ldots$.
As a result we have
\begin{eqnarray}
\label{intK7}
      K[r_{\mbox{st}}]&=&
     \exp\left(
     -\frac{1}{2}\Omega^{-1}\int d^4x_E\ \mbox{tr}\ 
     \ln (1-V^tLV)\right)
     \nonumber \\
     &=&
     \exp\left(
     -\frac{1}{2}\Omega^{-1}\int d^4x_E\ \mbox{tr}\ 
     \ln (1+F)\right),
\end{eqnarray} 
where 'tr' is to be understood as the trace in an ordinary matrix
sense. In the second equality we have used the property of matrix $V$, 
given by Eq.(\ref{matrixV}). This property can be used because of 
the trace before the logarithm. The matrix $F$ is equal to 
\begin{equation}        
            F_{\alpha\beta}=\frac{3\kappa}{16}A_{cbe}\left(
                   \begin{array}{cc}
                   -h^{(1)}_{ac}(s^e_{\mbox{st}}-h_e)&
                   h^{(1)}_{ac}p^e_{\mbox{st}}\\
                   h^{(2)}_{ac}p^e_{\mbox{st}}&
                   h^{(2)}_{ac}(s^e_{\mbox{st}}-h_e)
                   \end{array}
                   \right)_{\alpha\beta}.  
\end{equation}
The factor $\Omega^{-1}$ may be written as an ultraviolet divergent 
integral $\Omega^{-1}=\delta_E^4 (x-x)$. This singular term needs to 
be regularized, for instance, by introducing a cutoff $\Lambda_E$ damping 
the contributions from the large momenta $k_E$
\begin{equation}
   \Omega^{-1}=\delta_E^4 (0)
   \sim
   \int^{\Lambda_E /2}_{-\Lambda_E /2}\frac{d^4k_E}{(2\pi )^4} 
   =\frac{\Lambda_E^4}{(2\pi )^4}\ .
\end{equation}  

To finish our calculation one needs to return back to the Minkowski
space by replacing $x_4\rightarrow -ix_0$. It follows then that 
the functional integral (\ref{ancJ2}) is given by\footnote{The sign
of the term $\sim\Omega^{-1}$ must be corrected accordingly in 
Eq.(34) of \cite{Osipov:2002}.}
\begin{eqnarray} 
\label{Sr}
  &&{\cal Z}[\sigma ,\phi ;\Delta ]
    \sim e^{iS_r}, \\ 
  &&S_r=\int d^4x\left\{
    {\cal L}_r(r_{\mbox{st}})
   -\frac{\Omega^{-1}}{2}
        \sum_{n=1}^\infty\frac{(-1)^{n}}{n}\mbox{tr}
        \left[F^n_{\alpha\beta}(r_{\mbox{st}})\right]\right\}.
        \nonumber
\end{eqnarray}

The action (\ref{Sr}) contains in closed form all information about
$\hbar$-order corrections to the classical Lagrangian ${\cal 
L}_r(r_{\mbox{st}})$. Nevertheless it is still necessary to do some work to
prepare this result for applications. In the following we deal mainly
with the first term of the series, since it is the only one that 
contributes to the gap equation,
\begin{equation} 
\label{Sra} 
    {\cal L}_r={\cal L}_r(r_{\mbox{st}})+
      \frac{3\kappa}{32}\Omega^{-1}
      A_{abc}\left(
      h^{(2)}_{ab}-h^{(1)}_{ab}\right)(s_{\mbox{st}}^c-h_c)+\ldots
\end{equation}
Here one should sum over indices $a,b,c=0,1,\ldots 8$. In the next 
section we will calculate these sums for the cases with exact $SU(3)$
and broken $SU(3)\rightarrow SU(2)_I\times U(1)_Y$ flavor symmetry. 

To give some additional insight into the origin of formula (\ref{Sr}) 
let us note that $\hbar$-order corrections to 
${\cal L}_r(r_{\mbox{st}})$ can be obtained without evaluation of the 
Gaussian path integral in Eq.(\ref{ancJ2}). Instead one can start 
directly from Lagrangian (\ref{lam}) and obtain the one-loop
contribution using the canonical operator formalism of quantum field 
theory \cite{Bogoliubov:1980}. Since the Lagrangian does not contain 
kinetic terms the time-ordered products of meson fields have a pure 
singular form: 
$<T\{\sigma_a(x)\sigma_b(y)\}>=i\delta (x-y)h^{(1)-1}_{ab}\ ,\ 
<T\{\phi_a(x)\phi_b(y)\}>=i\delta (x-y)h^{(2)-1}_{ab}$. It is easy to
see, for instance, that the tadpole contribution, coming from the 
cubic terms in (\ref{lam}), exactly coincides with the term linear in
$\sigma$ in Eq.(\ref{Sra}).

%%%%%%%%%%%%%%%%%%%%%%%%%%%%%%%%%%%%%%%%%%%%%%%%%%%%%%%%%%%%%%%%%%%%%%%%%%%%%%%
\section{The ground state in the semiclassical expansion}
\label{sec-3}
%%%%%%%%%%%%%%%%%%%%%%%%%%%%%%%%%%%%%%%%%%%%%%%%%%%%%%%%%%%%%%%%%%%%%%%%%%%%%%%
The considered model belongs to the NJL-type models and therefore at
some values of coupling constants the many-fermion interactions can rearrange 
the vacuum into a chirally asymmetric phase, with mesons being the bound 
states of quark and antiquark pairs and unconfined quarks with reasonably large
effective masses. The process of the phase transition is governed by the gap 
equation, and, as we already know from Eq.(\ref{Sra}), the gap equation is 
modified by the additional contribution which comes from the term $\sim 
\Omega^{-1}$. Our aim now is to trace the consequences of this contribution 
on the ground state. 

An effective potential $V(\sigma_a)$ that describes the system in the 
chirally asymmetric phase has a minimum at some non-zero value of
$\sigma_a=\Delta_a$. The effective Lagrangian constructed at the 
bottom of this well does not contain linear terms in $\sigma$ fields. 
It means that the linear terms in Eq.(\ref{Sra}) must be canceled by 
the quark tadpole contribution. This requirement can be expressed in 
the following equation
\begin{eqnarray}
\label{gap}
   &&h_i+\frac{N_c}{2\pi^2}m_iJ_0(m_i^2)=z_i\ , 
   \nonumber \\
   &&z_i=\frac{3\kappa}{32}\Omega^{-1}
      \left(h^{(1)}_{ab}-h^{(2)}_{ab}\right)A_{abc}
      h^{(1)}_{ci}.
\end{eqnarray}
The relation of flavor indicies $i=u,d,s$ and indicies $a,b,c...$ is
given below in Eq.(\ref{eo}).
The first term, $h_i$, and the term on the right-hand side, $z_i$,
are the contributions from Lagrangian (\ref{Sra}). The second one is 
the contribution of the quark loop from (\ref{lagr2}) with a
regularized quadratically divergent integral $J_0(m^2)$ being defined as
\begin{eqnarray}
  J_0(m^2)&=&16\pi^2i\int_\Lambda\frac{d^4q}{(2\pi)^4}\frac{1}{q^2-m^2}
          =\int_0^\infty\frac{dt}{t^2}e^{-tm^2}\rho
           (t,\Lambda^2)
           \nonumber \\
          &=&\Lambda^2-m^2\ln\left(1+
           \frac{\Lambda^2}{m^2}\right).
\end{eqnarray}
This integral is a positive definite function for all real values of the 
cutoff parameter $\Lambda$ and the mass $m$. The kernel 
$\rho (t,\Lambda^2)=1-(1+t\Lambda^2)\exp (-t\Lambda^2)$ is
introduced through the Pauli -- Villars regularization of the integral 
over $t$, which otherwise would be divergent at the point $t=0$. Assuming
that $h_i$ is a known solution of Eqs.(\ref{saddle-1}), 
being a function of parameters $G,\ \kappa$ and the vacuum expectation
values $\Delta_i$ of scalar fields, we call Eq.(\ref{gap}) a gap equation. 
The right-hand side of this equation is suppressed by a factor $\hbar$ 
in comparison with the first two terms. Thus an exact solution to this 
equation, i.e. the constituent quark masses $m_i\ (i=u,d,s)$, would 
involve all powers of $\hbar$, but higher powers of $\hbar$ in such a 
solution would be affected by order $\hbar^2$ corrections to the 
equation. It is apparent that one has to restrict our solution only to 
the first two terms in an expansion of $m_i$ in powers of $\hbar$, 
obtaining $m_i$ in the form $m_i=M_i+\Delta m_i$. It is important 
to stress for physical applications that Eq.(\ref{gap}) gets at $\hbar$ 
order an additional contribution from meson loops 
\cite{Kleinert:2000,Babaev:2000} which we do not consider here. 
We restrict our attention only to the new kind of contribution of 
order $\hbar$ which is essential for the $N_f=3$ case and has not yet 
been discussed in the literature.

%%%%%%%%%%%%%%%%%%%%%%%%%%%%%%%%%%%%%%%%%%%%%%%%%%%%%%%%%%%%%%%%%%%%%%%

\subsection{Gap equation at leading order: $SU(3)$ case, 
            two minima}

%%%%%%%%%%%%%%%%%%%%%%%%%%%%%%%%%%%%%%%%%%%%%%%%%%%%%%%%%%%%%%%%%%%%%%%
The first two terms of Eq.(\ref{gap}) are of importance at leading 
order. Combining this approximation for Eq.(\ref{gap}) (i.e. setting 
$z_i\equiv 0$) together with Eq.(\ref{saddle-1}), one obtains the gap 
equation already known from a mean field approach 
\cite{Bernard:1988,Hatsuda:1994}, 
which self-consistently determines the constituent quark masses 
$M_i$ as functions of the current quark masses and coupling constants 
\begin{eqnarray}
\label{order-h0}
  \Delta_i&=&-Gh_i-\frac{\kappa}{32}\sum_{j,k}t_{ijk}h_jh_k,
            \nonumber \\
  h_i&=&-\frac{N_c}{2\pi^2}m_iJ_0(m_i^2)\equiv 2<\bar{q}_iq_i>\equiv
       2\alpha_i(m_i).
\end{eqnarray}  
Here the totally symmetric coefficients $t_{ijk}$ are equal to zero
except for the case with different values of indices $i\neq j\neq k$
when $t_{uds}=1$. The latin indices $i,j,k$ mark 
the flavour states $i=u,d,s$ which are linear combinations of states 
with indices $0,3$ and $8$. One projects one set to the other by the 
use of the following matrices $\omega_{ia}$ and $e_{ai}$ defined as 
\begin{equation}
\label{eo}
   e_{ai}=\frac{1}{2\sqrt 3}\left(
          \begin{array}{ccc}
          \sqrt 2&\sqrt 2&\sqrt 2 \\
          \sqrt 3&-\sqrt 3& 0 \\
          1&1&-2 
          \end{array} \right),\quad
   \!\!\omega_{ia}=\frac{1}{\sqrt 3}\left(
          \begin{array}{ccc}
          \sqrt 2&\sqrt 3& 1 \\
          \sqrt 2&-\sqrt 3& 1 \\
          \sqrt 2&0&-2 
          \end{array} \right)
\end{equation}
Here the index $a$ runs $a=0,3,8$ (for the other values of $a$ the 
corresponding matrix elements are assumed to be zero). We have then
$h_a=e_{ai}h_i$, and $h_i=\omega_{ia}h_a$. Similar relations can be 
obtained for $\Delta_i$ and $\Delta_a$. In accordance with this notation 
we use, for instance, that $h^{(1)}_{ci}=\omega_{ia}h^{(1)}_{ca}$.
The following properties of matrices (\ref{eo}) are straightforward:
$\omega_{ia}e_{aj}=\delta_{ij}, \ e_{ai}\omega_{ib}=\delta_{ab}$ 
and $e_{ai}e_{aj}=\delta_{ij}/2$. The coefficients $t_{ijk}$ are related 
to the coefficients $A_{abc}$ by the embedding formula 
$3\omega_{ia}A_{abc}e_{bj}e_{ck}=t_{ijk}$. The $SU(3)$ matrices 
$\lambda_a$ with index $i$ are defined in a slightly different
way $2\lambda_i=\omega_{ia}\lambda_a$ and $\lambda_a=2e_{ai}\lambda_i$.
In this case it follows that, for instance, $\sigma =\sigma_a\lambda_a=
\sigma_i\lambda_i=\mbox{diag}(\sigma_u,\sigma_d,\sigma_s)$, but 
$2\sigma_a\Delta_a=\sigma_i\Delta_i$.

It is well known \cite{Bernard:1988} that the gap equation at leading 
order has at least one nontrivial solution, $M_i(\hat{m}_j, G, \kappa , 
\Lambda )$, if $\kappa , \hat{m}_i\neq 0$. This important result is 
a direct consequence of the asymptotic behavior of the quark condensate 
$\alpha_i(m_i):\ \alpha_i(0)=\alpha_i(\infty )=0,\ \alpha_i(m_i)<0$ if 
$0<m_i<\infty$, from one side, and the monotonical decrease of the 
negative branch of the function $h_i(\Delta_j)$ on the semi-infinite 
interval $\hat{m}_i\leq m_i<\infty$, from the other side.
Stronger statements become possible if we have 
more information. Let us assume that $SU(3)$ flavour symmetry is preserved.
If $\hat{m}_u=\hat{m}_d=\hat{m}_s$, one can conclude that 
$\Delta_u=\Delta_d=\Delta_s$ and, as a consequence, we have
$h_u=h_d=h_s=\sqrt{2/3}h_0$. Instead of a system of three equations 
in (\ref{order-h0}) there is only one quadratic equation with two 
solutions 
\begin{eqnarray}
\label{u=d=s}
   h_u^{(1)}&=&-\frac{8G}{\kappa}\left(
       1-\sqrt{1-\frac{\kappa\Delta_u}{4G^2}}\right), 
       \nonumber \\
   h_u^{(2)}&=&-\frac{8G}{\kappa}\left(
       1+\sqrt{1-\frac{\kappa\Delta_u}{4G^2}}\right)\ .
\end{eqnarray} 
The second solution is always positive (for $\kappa <0, G>0$) and, due to 
this fact, cannot fulfill the second equation in (\ref{order-h0}). 
On the contrary, the solution $h_u^{(1)}$, as $\kappa\rightarrow 0$, 
gives $h_u^{(1)}=-\Delta_u/G$, which leads to the 
standard gap equation 

%%%%%%%%%%%%%%%%%%%%%%%%%%%    FIG.1    %%%%%%%%%%%%%%%%%%%%%%%%%%%%%%
\begin{figure}[t]
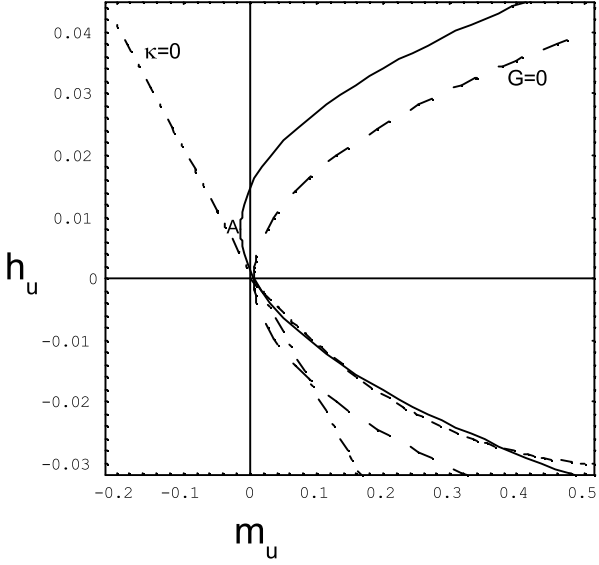

\PSfig{0.9}{fig.1}
\caption{The function $h_u(m_u)$ (here $[h_u]=\mbox{GeV}^3$, 
$[m_u]=\mbox{GeV})$ at fixed values 
$\hat{m}_u=6\ \mbox{MeV},\ \Lambda =860\ \mbox{MeV}$ for 
$G=0, \kappa =-5000\ \mbox{GeV}^{-5}$ (large dashes), 
$G=5\ \mbox{GeV}^{-2}, \kappa =0$ (dash-dotted line),
$G=5\ \mbox{GeV}^{-2}, \kappa =-5000\ \mbox{GeV}^{-5}$ (solid line). 
The small-dashes line corresponds to the function $2\alpha_u (m_u)$.}   
\label{fig1}
\end{figure}
%%%%%%%%%%%%%%%%%%%%%%%%%%%%%%%%%%%%%%%%%%%%%%%%%%%%%%%%%%%%%%%%%%%%%%%%
\noindent
$2\pi^2\Delta_i=N_cGm_iJ_0(m_i)$ for the theory 
without the 't Hooft determinant. Alternatively one could consider the 
theory with only the 't Hooft interaction (\ref{Ldet}) taking the limit 
$G\rightarrow 0$ in Eq.(\ref{u=d=s}). In this case we obtain for $h_u^{(1)}$
\begin{equation}
\label{Diak}
   h_u^{(1)}=\frac{4}{\kappa}\sqrt{-\kappa\Delta_u}
      =4\ \mbox{sign}(\kappa )\sqrt{\frac{\Delta_u}{-\kappa}}\quad (G=0).   
\end{equation} 

In Fig.\ref{fig1} we plot three different stationary trajectories 
$h_u=h_u(m_u)$ (with corresponding parameter values given in the caption) 
as functions of the quark mass $m_u$. To be definite
we put the current quark mass equal $\hat{m}_u=6\ \mbox{MeV}$ and 
the cutoff parameter is choosen to be $\Lambda =860\ \mbox{MeV}$. By fixing 
$\Lambda$ we completely fix the curve corresponding to the right-hand 
side of the gap equation, i.e. the function 
$2\alpha_u=-N_cm_uJ_0(m_u^2)/(2\pi^2)$. This function starts at the
origin of the coordinate system, being always negative for positive
values of $m_u$. At the point $m_u=\bar{m}\simeq 0.74\Lambda$ it has a 
minimum $\alpha_u ^{(min)}=-\Lambda^4\bar{m}/[3(\Lambda^2+\bar{m}^2)]$.
All stationary trajectories cross the $m_u$-axis 
at the point $m_u=\hat{m}_u$. The straight line corresponds to the
case $\kappa =0$. The second limiting case, $G=0$, is represented by
the solution (\ref{Diak}) and marked by $G=0$. 
The solid curve corresponds to Eq.(\ref{u=d=s}), starts at the turning 
point $A$ 
with the coordinates $A(m_u,h_u)=(\hat{m}_u+4G^2/\kappa , -8G/\kappa )$ 
and goes monotonically down $(h_u^{(1)})$ and up $(h_u^{(2)})$ for 
increasing values of $m_u$. 
The standard assignment of signs for the couplings $G$ and $\kappa$: 
$G>0, \kappa <0$ is assumed. The points where $h_u(m_u)$ intersects 
$2\alpha_u$ are the solutions of the gap equation. Let us remind 
\cite{Bernard:1988} that there are two qualitatively distinct classes 
of solutions. The first one is known as a solution with barely broken 
symmetry. We have this solution when $h_u$ intersects $2\alpha_u$ on 
the left side of its minimum (the minimum of $\alpha_u$ is outside the
region shown in the figure). If $h_u$ crosses $2\alpha_u$ on the right 
side of its minimum it corresponds to the case with the
firmly broken symmetry. It is known 
%%%%%%%%%%%%%%%%%%%%%%%%%%%    FIG.2    %%%%%%%%%%%%%%%%%%%%%%%%%%%%%%
\begin{figure}[t]
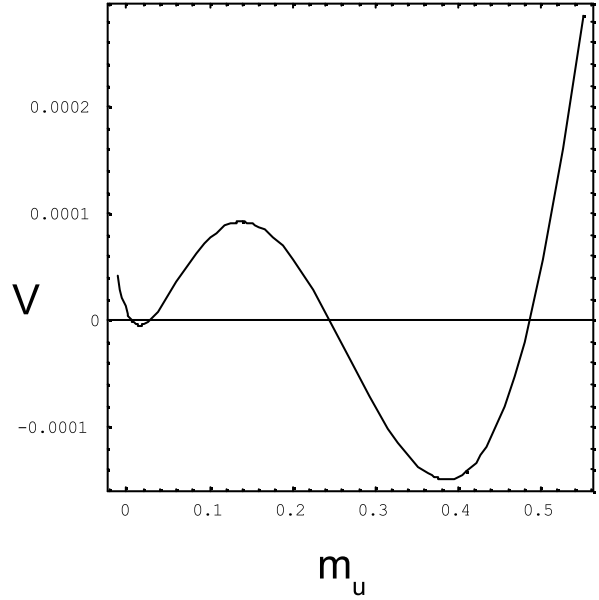

\PSfig{0.9}{fig.2}
\caption{The $\hbar$ leading order effective potential $V(m_u)$ 
         corresponding to the pattern shown by the solid line on 
         Fig.\ref{fig1}. We use the units $[V]=[m_u]=\mbox{GeV}$. }
\label{fig2}
\end{figure}
%%%%%%%%%%%%%%%%%%%%%%%%%%%%%%%%%%%%%%%%%%%%%%%%%%%%%%%%%%%%%%%%%%%%%%%%
\noindent
that the theory responds quite
differently to the introduction of a bare quark mass for these two
cases. The barely broken regime is characterized by strong 
non-linearities reflected in the behaviour of expectation values of
the scalar quark densities, $\bar{q}_iq_i$, in the physical quark 
states. Nevertheless, the solutions with the barely broken symmetry 
are likely to be more reliable from the physical point of view, in
particular, when $\Lambda\simeq 0.5\ \mbox{GeV}$ and higher. 

The six-quark interactions add several important new features into 
the picture. For instance, for some set of parameters, when 
$4G^2\sim\hat{m}_u|\kappa |$ or $G=0$, one can get three solutions 
of the gap equation, instead of one when $\kappa =0$. One of
these cases is illustrated in Fig.\ref{fig1} for the parameter set
$G=5\ \mbox{GeV}^{-2},\ \kappa =-5000\ \mbox{GeV}^{-5}$. 
The first solution is 
located quite close to the current quark mass value and, being a 
minimum of the effective potential, corresponds to the regime without 
the spontaneous breakdown of chiral symmetry. The next solution is a 
local maximum. The third one is a minimum and belongs to the regime with 
barely broken phase. The types of extrema are shown in Fig.\ref{fig2},
where we used Eq.(\ref{effpot}) with $\Omega^{-1}=0$ for the 
effective potential at leading order.  

The second new feature is that the stationary trajectories $h_u(m_u)$ 
in the case $\kappa\neq 0$ are real only starting from some value of 
$m_u\geq m_{min}$. For other values of $m_u$ where $m_u<m_{min}$ the 
effective energy is complex. To exclude this unphysical region the 
effective potential $V(m_u)$ must be defined as a single-valued
function on an half-open interval $m_u\geq m_{min}$.

%%%%%%%%%%%%%%%%%%%%%%%%%%%%%%%%%%%%%%%%%%%%%%%%%%%%%%%%%%%%%%%%%%%%%%%%%%% 

\subsection{NLO corrections: $z_i$ in the case of $SU(3)$ symmetry}

%%%%%%%%%%%%%%%%%%%%%%%%%%%%%%%%%%%%%%%%%%%%%%%%%%%%%%%%%%%%%%%%%%%%%%%%%%%
The functions $z_i$ can be evaluated explicitly. To this end it is 
better to start from the case with $SU(3)$ flavor symmetry. This 
assumption involves a significant simplification in the structure of 
the semiclassical corrections, giving us however a result which 
possesses all the essential features of the more elaborate cases.
Noticing that as a consequence of $SU(3)$ symmetry the function $h_a$
has only one non zero component $h_0$, one can find
\begin{equation}
\label{block}
   A_{acb}h_b=A_{ac0}h_0=\frac{1}{3}\left(2\delta_{a0}\delta_{c0}
              -\delta_{a\alpha}\delta_{c\alpha}\right)h_u
\end{equation}
where $\alpha =1,2\ldots 8.$  
This result determines in general the structure of all flavor 
multi-indices objects like, for instance, $h_{ab}^{(1,2)}$ defined by the 
Eq.(\ref{ha}). Indeed, taking into account (\ref{block}), one can represent 
algebraic equations for $h_{ab}^{(1,2)}$ in the form
\begin{eqnarray}
  &&G\left[\delta_{a0}\delta_{c0}
         \left(1\pm 2\omega\right) 
        +\delta_{a\alpha}\delta_{c\alpha}
         \left(1\mp\omega\right)\right]
         h_{cb}^{(1,2)}=-\delta_{ab}\ , 
       \nonumber \\
  &&  \omega =\frac{\kappa h_u}{16G}   
\end{eqnarray} 
where we follow the notation explained in Appendix \ref{A1}.
Since this expression is a diagonal matrix, one can easily obtain
\begin{equation}
\label{h_cb1}
   h_{cb}^{(1,2)}=-\frac{\delta_{c0}\delta_{b0}}{\displaystyle
          G\left(1\pm 2\omega\right)} 
          -\frac{\delta_{c\alpha}\delta_{b\alpha}}{\displaystyle
          G\left(1\mp\omega\right)}\ ,
\end{equation}
with $h_{cb}^{(1)}$ associated with the upper sign.

Consider now the perturbation term of Eq.(\ref{gap}). One can verify
that
\begin{eqnarray}
\label{differ}
   &&(h_{ab}^{(1)}-h_{ab}^{(2)})A_{abc}=\frac{\kappa h_u}{8G^2}\left(
   \frac{2A_{00c}}{1-4\omega^2}-\frac{\displaystyle 
   \sum_{\alpha=1}^{8}A_{\alpha\alpha c}}{1-\omega^2}\right)
       \nonumber \\
   &=&
   \frac{\kappa h_u}{2G^2}\sqrt{\frac{2}{3}}\delta_{c0}
   \frac{1-3\omega^2}{(1-4\omega^2)(1-\omega^2)}\ .
\end{eqnarray}
Here we used the properties of the coefficients $A_{abc}$ 
\begin{equation}
   A_{00c}=\frac{2}{3}\sqrt{\frac{2}{3}}\delta_{c0}, \qquad
   A_{abb}\equiv \sum_{b=0}^{8}A_{abb}=-2\sqrt{\frac{2}{3}}
   \delta_{a0}.
\end{equation}         
The quantum effect of auxiliary fields to the gap equation stems from 
the term
\begin{equation}
\label{dif}
   (h_{ab}^{(1)}-h_{ab}^{(2)})A_{abc}h_{ci}^{(1)} 
   =-\frac{\kappa h_u}{3G^3}
   \frac{1-3\omega^2}{(1+2\omega )(1-4\omega^2)(1-\omega^2)}
\end{equation}
where the right-hand side is the same for the three possible choices of
the index $i=u,d,s$. We conclude that in the $SU(3)$ limit the
functions $z_i$ are uniquely determined by  
\begin{equation}
\label{zsu3}
   z_u=-\frac{\kappa}{2G^2}\Omega^{-1}\frac{\omega 
       (1-3\omega^2)}{(1+2\omega )(1-4\omega^2)(1-\omega^2)}\ .
\end{equation}

A direct application of (\ref{zsu3}) is the evaluation of the
corrections $\Delta m_i$ to the constituent quark mass, 
\begin{equation}
\label{corr}
  \Delta m_u=\frac{\kappa}{2G}\Omega^{-1}
  \frac{\omega (1-3\omega^2)}{(1-4\omega^2)(1-\omega^2)}\
  {\cal Q}_{u}\ . 
\end{equation}
This result is based on Eq.(\ref{dmsu3}) and Appendix \ref{A2}. 
The right-hand side 
must be calculated with the leading order value of $m_u=M_u$. 
The factor ${\cal Q}_{u}$ is 
\begin{equation}
\label{omegauu}
  {\cal Q}_{u}=\frac{1}{1-(1+2\omega)I_u(m_u^2)}\ .
\end{equation}
Here the function $I_i(m_i^2)$ is proportional to the derivative of the 
quark condensate $<\bar{q}_iq_i>$ with respect to $m_i$ 
\begin{equation}
\label{I_i}
   I_i\equiv\frac{N_cG}{2\pi^2}\left[J_0(m_i^2)-2m_i^2J_1(m_i^2)\right],
   \qquad i=u,d,s
\end{equation}  
where the integral $J_1$ is given by
\begin{eqnarray}
  J_1(m^2)&=&-\frac{\partial}{\partial m^2}J_0(m^2)
          =\int_0^\infty\frac{dt}{t}e^{-tm^2}\rho
           (t,\Lambda^2)
                 \nonumber \\
          &=&\ln\left(1+\frac{\Lambda^2}{m^2}\right)
           -\frac{\Lambda^2}{\Lambda^2+m^2}\ .
\end{eqnarray} 

One can show that ${\cal Q}_{u}$ is related to the expectation values 
of the scalar quark density, $\bar{q}_iq_i$ in the physical quark
state $|Q_u>$. To be precise we have ${\cal Q}_u={\cal Q}_{uu}+{\cal Q}_{ud}
+{\cal Q}_{us}$, where the expectation values ${\cal Q}_{ui}$ are given 
in Appendix \ref{A2}.
It is of interest to know the sign of the quasi-classical correction 
$\Delta m_u$. In general the answer on this question depends on the 
values of coupling constants which should be fixed from the hadron mass
spectrum. All what we know at the moment is only that $G>0,\ \kappa
<0,\ \hat{m}_u\simeq 6\ \mbox{MeV}$ and $\Lambda\sim 1\ \mbox{GeV}$. 
One can expect also that the dynamical masses of the quarks $M_u$ are 
close to their empirical value $M_u\simeq M_N/3\sim 300\ 
\mbox{MeV}$, with $M_N$ the nucleon mass. Let us suppose now that
$0<\omega\ll 1$, what actually means that the coupling constants
belong to the interval $0<-\kappa (M_u-\hat{m}_u)/(4G)^2\ll 1$. 
This range is preferable from the point of view of $1/N_c$ counting. 
In this case we have
\begin{equation}
  \Delta m_u\simeq \frac{\kappa\omega}{2G}\ \Omega^{-1}
  {\cal Q}_{u}^{\omega =0} 
\end{equation}
and one can conclude that the sign of $\Delta m_u$ is opposite to the 
sign of ${\cal Q}_{u}^{\omega =0}$. In turn the function 
${\cal Q}_{u}^{\omega =0}$ is positive in some physically preferable 
range of values of $G$ such that $G\sim 5\ \mbox{GeV}^{-2}$. 

It must be emphasized that the approximation made in Eq.(\ref{corr}) is
legitimate only if the quasi-classical correction $\Delta m_u$ is small 
compared with the leading order result $M_u$. In particular, it is
clearly inapplicable near points where the function ${\cal Q}_{u}(m_u)$ 
has a pole. One sees from Eq.(\ref{omegauu}) that this takes place 
beyond some large values of $G$ or $\kappa$. There is a set of parameters 
for which the function $(1+2\omega )I_u(m_u^2)$ is always less than $1$. 
This is the case, for instance, for the choice just considered above.  
For large couplings, ${\cal Q}_{u}$ may have a pole, and one has 
to check that the mean field result $M_u$ is located at a safe
distance from them before using formula (\ref{corr}). 
The large couplings contain also the 
potential danger to meet the poles at the points $\omega =1$ and
$\omega =1/2$. These poles are induced by caustics in the Gaussian 
path integral and occur as singularities in the effective potential. 

We conclude this section with the expression for the quark condensate
at next to leading order in $\hbar$,
\begin{equation}
\label{hcortoqc}
    <\bar{q}_uq_u>=<\bar{q}_uq_u>_{0}
    -\frac{\Delta m_u}{2G}I_u(M_u^2),
\end{equation}
where the subscript $0$ denotes that the expectation value has been 
obtained in the mean field approximation.

%%%%%%%%%%%%%%%%%%%%%%%%%%%%%%%%%%%%%%%%%%%%%%%%%%%%%%%%%%%%%%%%%%%%%%%%% 
\subsection{NLO contribution to the effective potential: 
            $SU(3)$ symmetric result}
\label{effpot1}
%%%%%%%%%%%%%%%%%%%%%%%%%%%%%%%%%%%%%%%%%%%%%%%%%%%%%%%%%%%%%%%%%%%%%%%% 
The expectation value of the energy density in a state for which the 
scalar field has the expectation value $\Delta_i$ is given by the 
effective potential $V(\Delta_i)$. The effective potential is a direct
way to study the ground state of the theory. If $V(\Delta_i)$ has
several local minima, it is only the absolute minimum that corresponds 
to the true vacuum. A sensible approximation method to calculate 
$V(\Delta_i)$ is the semiclassical expansion \cite{Coleman:1988}. 
The first term in the expansion of $V$ is the classical potential. In 
the considered theory it contains the negative sum of all nonderivative 
terms in the bosonized Lagrange density which includes the one-loop 
quark diagrams and the leading order SPA result (\ref{lam}). 
The second term contains semiclassical corrections from Eq.(\ref{Sr}) 
and the one-loop meson diagrams. However, one can obtain the effective
potential directly from the gap equation. Indeed, let us assume that 
the potential $U(\sigma_a,\phi_a)$ of the Lagrange density of the 
bosonized theory is known, then $<U(\sigma_a,\phi_a)>=V(\Delta_i)$. 
To explore the properties of the spontaneously broken theory, we restrict
ourselves to the part of the total potential, $U(\sigma_i)$,
involving only the fields which develop a nonzero vacuum expectation 
value, $<\sigma_i>=\Delta_i$. Expanding $U(\sigma_i)$ about the 
asymmetric ground state, we find
\begin{equation}
\label{Uexp}
  U(\sigma_i)=U(\Delta_i)
    +\left.\frac{\partial U}{\partial\sigma_i}
    \right|_{\sigma_i=\Delta_i}(\sigma_i-\Delta_i)
    +\ldots\ .  
\end{equation}
It is clear that $U(\Delta_i)=V(\Delta_i)$, and the derivatives are 
functions of $\Delta_j$ 
\begin{equation}
\label{der}
    \left.\frac{\partial U}{\partial\sigma_i}\right|_{\sigma_i=\Delta_i}
    =\left\langle \frac{\partial U}{\partial\sigma_i}\right\rangle 
    =\frac{\partial V(\Delta_j)}{\partial\Delta_i}=f_i(\Delta_j).    
\end{equation}
It means, in particular, that we can consider Eq.(\ref{der}) as a 
system of linear differential equations to extract the effective 
potential $V(\Delta_i)$, if the dependence $f_i(\Delta_j)$ is known.

%%%%%%%%%%%%%%%%%%%%%%%%%%%    FIG.3    %%%%%%%%%%%%%%%%%%%%%%%%%%%%%%
\begin{figure}[t]
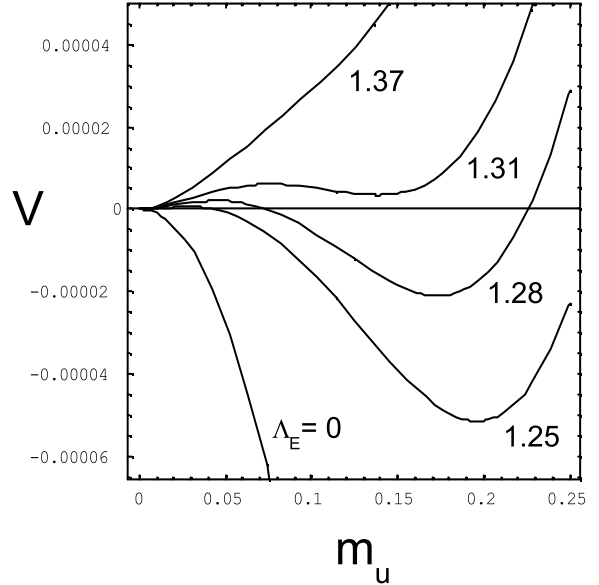

\PSfig{0.9}{fig.3}
\caption{The effective potential $V(m_u)$ to $\hbar$-order
         corresponding to the case $\kappa =-1800\ \mbox{GeV}^{-5}$,
         $G=10\ \mbox{GeV}^{-2}$, $\hat{m}=0$ and $\Lambda=860\ 
         \mbox{MeV}$, where the values of $\Lambda_E$ are a measure
         for the strength of the NLO corrections in the effective 
         potential. The units are $[V]=[m_u]=\mbox{GeV}$.} 
\label{fig3}
\end{figure}
%%%%%%%%%%%%%%%%%%%%%%%%%%%%%%%%%%%%%%%%%%%%%%%%%%%%%%%%%%%%%%%%%%%%%%%%

Further, Eq.(\ref{Uexp}) tells us that $f_i(\Delta_j)$ are determined 
by the tadpole term in a shifted potential energy,
$U(\sigma_i'+\Delta_i)$, where we define a new quantum field with 
vanishing vacuum expectation value $\sigma_i'=\sigma_i-\Delta_i$. 
In the case of $SU(3)$ flavor symmetry, we have $f_i=f$ where 
$i=u,d,s$ and $f$ is given by
\begin{eqnarray}
\label{f}
   f(m_u)&=&-\frac{h_u}{2}-\frac{N_c}{4\pi^2}m_uJ_0(m_u^2)
             \nonumber \\
         &-& \frac{\kappa\Omega^{-1}
             \omega (1-3\omega^2)}{4G^2(1+2\omega )
             (1-4\omega^2)(1-\omega^2)}\
   .
\end{eqnarray}  
Therefore, the condition for the extremum $\partial V(\Delta_u 
)/\partial\Delta_u =0$ coincides with the gap equation (\ref{gap}). 
In Appendix \ref{A3} we obtain from Eq.(\ref{der}) in the 
$SU(3)$ limit the effective potential
\begin{eqnarray}
\label{effpot}
  V(\Delta_u)&=&\frac{1}{4}\left(3Gh_u^2+\frac{\kappa}{4}h_u^3\right)
             -\frac{3}{2}v(m_u^2)
              \nonumber \\
             &-&\frac{1}{2}\Omega^{-1}\ln\left|
               (1-4\omega^2)(1-\omega^2)^8\right|-C.
\end{eqnarray}
The free constant $C$ can be fixed by requiring $V(0)=0$. In this
expression $v(m_u^2)$ is defined according to Eq.(\ref{v}) and
$h_u$ is the first of the two solutions of the stationary point equation 
given by (\ref{u=d=s}). Notice, that these solutions 
are complex when $4G^2<\kappa\Delta_u$. Hence the function $V(\Delta_u)$
is not real as soon as the inequality is fulfilled. The most efficient 
way to go round this problem and to define the effective potential as a 
real function on the whole real axis is to treat $h_u$ as an independent 
variational parameter instead of $\Delta_u$ in Eq.(\ref{effpot}). 
In this approach, which actually corresponds more closely to the BCS 
theory of superconductivity, one should consider equation 
(\ref{sph}) as the one which yields the function $\Delta_u(h_u)$. 
One can check now that the extremum condition $\partial 
V(h_u)/\partial h_u=0$ is equivalent to the gap equation (\ref{gap}) 
where the quark mass is expressed in terms of $h_u$ \cite{Hatsuda:1994}. 
Thus, the effective potential in the form of $V(h_u)$ provides for a 
direct way to determine the minimum of the vacuum energy
irrespectively of the not well defined mapping $\Delta_u\rightarrow h_u$. 
 
There is a direct physical interpretation of Eq.(\ref{effpot}): 
classically, the system sits in a minimum of the potential energy,
$U_{cl}$, determined by the first two terms, and its energy is the
value of the potential at the minimum, $U_{cl}(M_u)$. To get the 
first quantum correction, $\Delta U$, to this picture, we add 
the third term $(\sim\Omega^{-1})$, and approximate the potential,
$V(M_u+\Delta m_u)$, near the classical minimum by a function 
$U_{cl}(M_u)+\Delta U(M_u)$.

In Fig.\ref{fig3} we show the effective potential calculated for
$\kappa =-1800\ \mbox{GeV}^{-5},\  G=10\ \mbox{GeV}^{-2},\ \hat{m}=0$
and $\Lambda =860\ \mbox{MeV}$ depending on the strength of the 
fluctuations, indicated by the Euclidean cutoff $\Lambda_E$. In
absence of fluctuations the minimum occurs at $m_u=M_{min}=340\ 
\mbox{MeV}$ (outside the range of this figure). Increasing the effect 
of fluctuations, the minima $M_{min}$ appear at smaller values 
and the potential gets shallower. Simultaneously a barrier develops
between $V(0)$ and $V(M_{min})$. At some critical value of
$\Lambda_E$ the point $V(0)$ becomes the stable minimum and the trivial
vacuum is restored\footnote{Away from the chiral limit the barier 
between the two vacua ceases fast to exist and the transition from
one phase to the other occurs smoothly.}.
This effect has a simple explanation. 
In the neighbourhood of the trivial vacuum where $m_u$ is small
the effective potential $V(m_u)$ can be well described by the first  
terms of the series in powers of $m_u$
\begin{eqnarray}
   \left. V(\Delta_u)\right|_{\hat m=0}&=&
   \frac{3m_u^2}{4G}\left[
   1-\frac{N_cG\Lambda^2}{2\pi^2}+\frac{\kappa^2}{32G^3}
     \left(\frac{\Lambda_E}{2\pi}\right)^4\right]
     \nonumber \\
    &+&{\cal O}(m_u^3).
\end{eqnarray}
The trivial vacuum always exists when 
\begin{equation}
\label{trivvac}
    1-\frac{N_cG\Lambda^2}{2\pi^2}+\frac{\kappa^2}{32G^3}
     \left(\frac{\Lambda_E}{2\pi}\right)^4\ge 0.
\end{equation} 
This inequality generalizes the well-known result for $\kappa =0$.

The local minima $M_{min}$ in the broken phase are the {\it exact}
solutions of the full gap equation (\ref{gap}). We find that at leading 
$\hbar$-order, $M_{min}^{pert}=M_u+\Delta m_u$, where $\Delta m_u$
is the correction (\ref{corr}), follows within a few 
percent the pattern shown for $M_{min}$ in Fig.\ref{fig3}.
For instance, one has at $\Lambda_E=1.25\ \mbox{GeV},\  
M_{min}^{pert}=214\ \mbox{MeV}$, and at $\Lambda_E=1.31\ \mbox{GeV},\ 
M_{min}^{pert} = 188\ \mbox{MeV}$. 
It is clear that the phase transition shown in Fig.\ref{fig3} is a 
non-perturbative effect. Instead the perturbative result yields 
$M_{min}^{pert}\rightarrow 0$ smoothly with increasing $\Lambda_E$
up to the value $\Lambda_E\simeq 1.6$ GeV. 

Going to higher values of $m_u$ (not shown in Fig.\ref{fig3}) one can 
come to caustics, i.e. singularities in $V(\Delta_u)$. From the 
logarithm in Eq.(\ref{effpot}) we obtain the values $\Delta_u$ where
it happens. There are two singular points 
\begin{equation}
   \Delta_u^{(1)}=-12\frac{G^2}{\kappa}\ , \qquad 
   \Delta_u^{(2)}=-32\frac{G^2}{\kappa}\ .
\end{equation}
For given values of couplings $G$ and $\kappa$ we have 
$\Delta_u^{(1)}=667\ \mbox{MeV}$ and $\Delta_u^{(2)}=1.78\ \mbox{GeV}$. 
The indicated curves with $\Lambda_E\neq 0$ have as asymptote the 
vertical line crossing the $m_u$-axis at the point $\Delta_u^{(1)}$.
It is clear that for other parameter choices the ordering 
$\Delta_u^{(1)}<M_u<\Delta_u^{(2)}$, or even 
$\Delta_u^{(1)}<\Delta_u^{(2)}<M_u$ are possible, where $M_u$ is 
the classical minimum. In these cases a careful 
treatment of the caustic regions must be done.

%%%%%%%%%%%%%%%%%%%%%%%%%%%%%%%%%%%%%%%%%%%%%%%%%%%%%%%%%%%%%%%%%%%%%%%%%%%%
\subsection{Gap equation at leading order:
            $SU(2)_I\times U(1)_Y$ case, general properties}
%%%%%%%%%%%%%%%%%%%%%%%%%%%%%%%%%%%%%%%%%%%%%%%%%%%%%%%%%%%%%%%%%%%%%%%%%%%%
We will now apply the same strategy to the case $\hat{m}_u=\hat{m}_d
\neq\hat{m}_s$, which breaks the unitary $SU(3)$ symmetry down to the
$SU(2)_I\times U(1)_Y$ (isospin-hypercharge) subgroup. 
In full agreement with symmetry requirements it follows 
then that $m_u=m_d\neq m_s$ and $h_u=h_d\neq h_s$. Thus, we have a 
system of two equations (from Eq.(\ref{saddle-1})) to determine the
functions $h_u$ and $h_s$. These equations can be easily solved in the 
limit $G\rightarrow 0$
\begin{equation}
\label{Diak-2}   
   h_u=-4\sqrt{\frac{\Delta_s}{-\kappa}}\ \ ,\qquad
   h_s=\frac{4}{\kappa}\Delta_u\sqrt{\frac{-\kappa}{\Delta_s}}
       \quad (G=0).   
\end{equation} 
Obviously the result Eq.(\ref{Diak}) follows from these expressions.    
For $G\neq 0$ the system is equivalent to the following one     

%%%%%%%%%%%%%%%%   FIG.4   %%%%%%%%%%%%%%%%%%%%%%%%%%%%%%%%%%%%%%%%%%%%%%%%%
\begin{figure}[t]
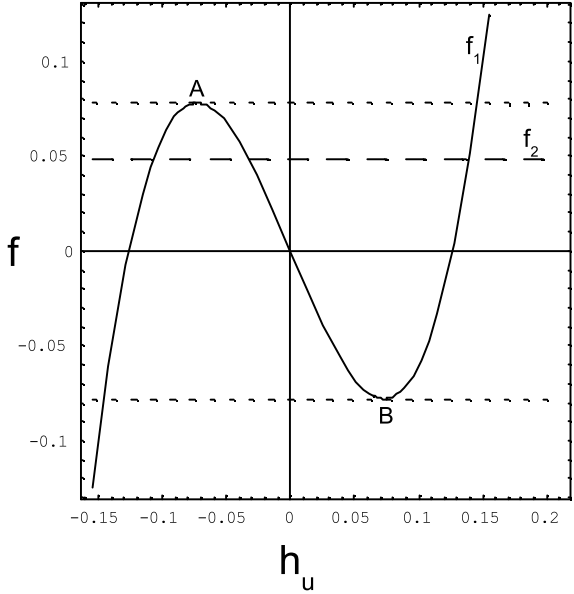

\PSfig{0.9}{fig.4}
\caption{Graphical solution of the cubic equation in
(\ref{huhs}). The left-hand $(f_1)$ and the right-hand $(f_2)$ sides
of this equation are plotted as a function of $h_u$ for $x_s<1$. The region
bounded by the dashed lines corresponds to the interval $D<0$ where 
the cubic equation has three real solutions. The local maximum $A$ and the 
local minimum $B$ have coordinates $A: \{h_u=-\sqrt{-Q},\ f=\Delta
(x_s)/G\}$ and $B: \{h_u=\sqrt{-Q},\ f=-\Delta (x_s)/G\}$.} 
\label{fig4}
\end{figure}
%%%%%%%%%%%%%%%%%%%%%%%%%%%%%%%%%%%%%%%%%%%%%%%%%%%%%%%%%%%%%%%%%%%%%%%%%%%%%
\begin{equation}
\label{huhs} 
   \left\{
   \begin{array}{rcl}
   \displaystyle
   \left(\frac{\kappa}{16G}\right)^2h_u^3+(x_s-1)h_u
   &=&\displaystyle\frac{\Delta_u}{G}\\
   && \\
   Gh_s+\Delta_s+\displaystyle\frac{\kappa}{16}h_u^2&=&0\\
   \end{array}
   \right.
\end{equation}
where we put $x_s=\kappa\Delta_s/(4G)^2$ in accordance with our 
notation in the Appendix \ref{A1}. From the physics of 
instantons we know that the strength constant $\kappa <0$. It makes 
$x_s$ negative. Hence the left-hand side of the first equation is not 
a monotonic function of $h_u$ (see Fig.\ref{fig4}). The right-hand side is a 
positive constant, if $\Delta_u>0$ and $G>0$, what is usually assumed. 
Therefore, the equation has three different real solutions, 
$h_u^{(n)}(\Delta_u, \Delta_s),\ n=1,2,3$, in some interval of values 
for $\Delta_u/G$. The boundaries of the interval are given by the 
inequality $-\Delta (x_s)<\Delta_u<\Delta (x_s)$, where
\begin{equation}
   \Delta (x_s)=\frac{32G^2}{|\kappa |}\left(
                \frac{1-x_s}{3}\right)^{3/2}_. 
\end{equation}

By means of the discriminant $D$ of the cubic equation, $D=Q^3+R^2$, where
\begin{equation}
   Q=\left(\frac{16G}{\kappa}\right)^2\frac{x_s-1}{3}\ , \qquad
   R=\left(\frac{16G}{\kappa}\right)^3\frac{x_u}{2}\ ,
\end{equation}

%%%%%%%%%%%%%%%%   FIG.5   %%%%%%%%%%%%%%%%%%%%%%%%%%%%%%%%%%%%%%%%%%%%%%%%%
\begin{figure}[t]
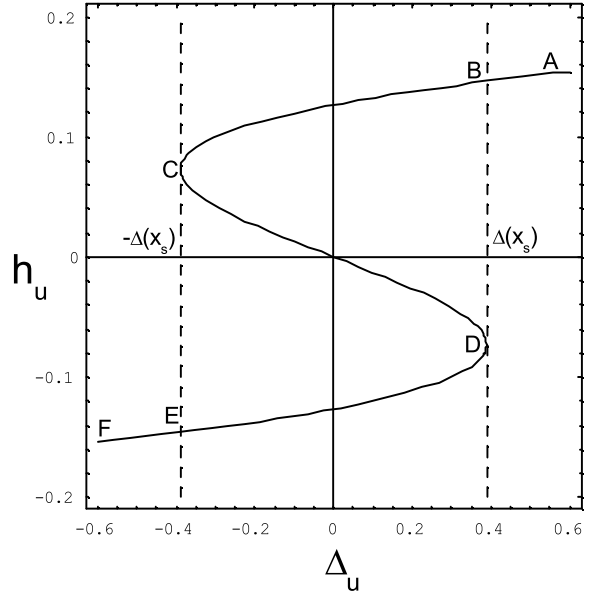

\PSfig{0.9}{fig.5}
\caption{The stationary trajectory $h_u(\Delta_u)$ at 
fixed $\Delta_s$. Inside the interval $|\Delta_u|<\Delta (x_s)$
it is a multi-valued mapping. The monotonic curves: $BC, CD, DE$
correspond to the three well-defined single-valued functions: 
$h_u^{(n)}, n=1,2,3.$} 
\label{fig5}
\end{figure}
%%%%%%%%%%%%%%%%%%%%%%%%%%%%%%%%%%%%%%%%%%%%%%%%%%%%%%%%%%%%%%%%%%%%%%%%%%%%%
\noindent
this region can be shortly identified by $D<0$. The qualitative picture 
of the dependence $h_u(\Delta_u)$ at a fixed value of $\Delta_s$ is shown 
in Fig.\ref{fig5}. The three solutions in the region $D\le 0$ can be 
parametrized by the angle $\varphi$
\begin{eqnarray}
\label{huGneq0}
  h_u^{(1)}&=&2\sqrt{-Q}\cos\frac{\varphi}{3}\ , \quad 
  h_u^{(2)}=2\sqrt{-Q}\sin\left(\frac{\varphi}{3}-\frac{\pi}{6}\right),
  \nonumber \\
  h_u^{(3)}&=&-2\sqrt{-Q}\sin\left(\frac{\varphi}{3}+\frac{\pi}{6}\right) 
\end{eqnarray} 
where
\begin{equation}
\label{cosphi}
   \cos\varphi =\frac{R}{\sqrt{-Q^3}}\ , \quad 
   \sin\varphi =\sqrt{1+\frac{R^2}{Q^3}}\ .
\end{equation}
The angle $\varphi$ can always be converted to values of $\varphi$ 
such that $0\le\varphi\le\pi$. The boundaries $\varphi =0$ and
$\varphi =\pi$ correspond to the value $D=0$. When the argument 
$\varphi$ increases from $0$ to $\pi$, the solutions $h_u^{(1)}, 
h_u^{(2)}$ and $h_u^{(3)}$ run along the curves $BC, DC$ and $DE$ 
accordingly. These curves intersect the $h_u$-axis at $\varphi
=\pi/2$, where $R=0$. One can show that $h_u^{(2)}\rightarrow 
-\Delta_u/G$ at $\kappa\rightarrow 0$ in full agreement with our 
previous result following from Eq.(\ref{u=d=s}). On 
the other side $h_u^{(3)}$ leads to the result (\ref{Diak-2}) in the 
limit $G\rightarrow 0$. Previously, studying the $SU(3)$ case, we
have obtained both of these limits from one solution (\ref{u=d=s}).
Now one has to use either solution $h_u^{(2)}$ or solution 
$h_u^{(3)}$ depending on the values of the parameters
$\kappa , G, \Delta_u$, and $\Delta_s$ which correspond to the minimum
of the energy density. There is no chance to join the 
partial cases $G=0$ and $\kappa =0$ in one solution, because they lead 
to systems of quadratic (in the first case) and linear (in the second
case) equations without intersection of their roots. The stationary
trajectory $h_u^{(1)}$ is a positive definite function and thus the
solution of the gap equation does not belong to this branch. 

Let us suppouse that the system of two equations which describe the
vacuum state of the theory in the case of the $SU(2)_I\times U(1)_Y$
symmetry at leading order has a solution, i.e., the constituent quark 
masses $M_{u(1)}$ and $M_{s(1)}$, corresponding to a local extremum of 
the effective potential $V(m_u, m_s)$, are known. We may assume that 
$M_{u(1)}$ belongs to the region with the barely broken symmetry, 
$M_{u(1)}<\bar{m}_u$. As we already know, it is the most preferable 
pattern from the physical point of view. Then there is no any other 
solution $M_{u(1)}'\neq M_{u(1)}$ for this set of parameters $G, \kappa , 
\Lambda , \hat{m}_u, \hat{m}_s$ and already fixed value $M_{s(1)}$. 
This follows from the pure geometrical fact that the second order 
derivatives for curves $h_u(\Delta_u)$ and $<\bar{q}_uq_u>$ have 
opposite signs in the considered region. 
Suppose further that there are other
solutions $M_{u(n)},\ M_{s(n)}$ with $n=2,3.$ This statement does 
not contradict our previous result corresponding to the case 
with the $SU(3)$ symmetry, where we were able to find out three 
solutions for some sets of parameters. The functions $M_{u(n)}$ and 
$M_{s(n)}$ can be understood as chiral expansions about the
$SU(3)$ symmetric solutions. The coefficients of the chiral series  
are determined by the expectation values ${\cal Q}_{ij}(M_u)$ and their
derivatives. It means that solutions in the case of $SU(3)$
symmetry are related to the solutions for the more general 
$SU(2)_I\times U(1)_Y$ case. Hence, there are only three sets of solutions 
$(M_{u(n)}, M_{s(n)})$ at maximum (for the considered region). The
effective potential helps us to classify these critical points as
will be discussed in Sec. \ref{ssF}.

%%%%%%%%%%%%%%%%%%%%%%%%%%%%%%%%%%%%%%%%%%%%%%%%%%%%%%%%%%%%%%%%%%%%%%%%%%%%% 

\subsection{NLO corrections: $z_i$ and $\Delta m_i$ in the case of 
            $SU(2)_I\times U(1)_Y$ symmetry}

%%%%%%%%%%%%%%%%%%%%%%%%%%%%%%%%%%%%%%%%%%%%%%%%%%%%%%%%%%%%%%%%%%%%%%%%%%%%%
Our next task is to take into account quantum fluctuations and compute
the corresponding corrections to the leading order result $M_u, M_s$.
For this purpose we need to find $z_u$ and $z_s$.
Consider first the sum $A_{acb}h_b$ which can be written as a $9\times 
9$ matrix in block diagonal form 
\begin{equation}
   A_{acb}h_b=-\frac{1}{3}\left(
              \begin{array}{ccc}
              \Omega_1 & 0 & 0 \\
              0 & \Omega_2 & 0 \\
              0 & 0 & \Omega_3
              \end{array}
              \right)
\end{equation}
with $2\times 2$, $3\times 3$ and $4\times 4$ blocks
\begin{eqnarray}
   (\Omega_1)_{rs}&=&\frac{1}{3}\left(
            \begin{array}{cc}
            -2(2h_u+h_s) & \sqrt{2}(h_u-h_s) \\
            \sqrt{2}(h_u-h_s) & (4h_u-h_s)
            \end{array}
            \right)_{rs}\ , 
            \nonumber \\
   (\Omega_2)_{nm}&=&h_s\delta_{nm}\ , \quad
   (\Omega_3)_{fg}=h_u\delta_{fg}\ .           
\end{eqnarray}
The indicies $r,s$ of the first matrix range over the subset $r,s=0,8$ 
of the set $a,c=0,1,\ldots 8$. In the matrix $\Omega_2$ we assume that 
$n,m=1,2,3$ and in $\Omega_3$ the indices take values $f,g=4,5,6,7$. 

Using this result one can solve the last two equations in (\ref{ha}), 
rewriting them in the form
\begin{equation}
   G\left(
          \begin{array}{ccc}
          1\mp\displaystyle\frac{\kappa\Omega_1}{16G} & 0 & 0 \\
          0 & 1\mp\displaystyle\frac{\kappa\Omega_2}{16G} & 0 \\
          0 & 0 & 1\mp\displaystyle\frac{\kappa\Omega_3}{16G}
          \end{array} 
    \right)_{ac}\!\!h_{ce}^{(1,2)}=-\delta_{ae}
\end{equation}
and find the functions $h_{ce}^{(1,2)}$. We obtain
\begin{equation}
   h_{nm}^{(1,2)}=\frac{-\delta_{nm}}{G(1\mp\omega_s)}\ , \qquad
   h_{fg}^{(1,2)}=\frac{-\delta_{fg}}{G(1\mp\omega_u)}\ , 
\end{equation}
where $\omega_i$ are defined in the Appendix \ref{A1}. For the 
$2\times 2$ matrix with indices $0,8$ we have
\begin{eqnarray}
   h^{(1,2)}_{rs}&=&\frac{-1}{3G(1\pm\omega_s-2\omega_u^2)}
          \nonumber \\
       &&\left(
         \begin{array}{cc}
         3\mp (4\omega_u-\omega_s) & \
         \pm\sqrt{2}(\omega_u-\omega_s) \\
     \\  \pm\sqrt{2}(\omega_u-\omega_s) & \
         3\pm 2(2\omega_u+\omega_s)  
         \end{array}
   \right)_{rs}. 
\end{eqnarray}
In particular, if the terms $\omega_u$ and $\omega_s$ are equal,
these expressions coincide with Eq.(\ref{h_cb1}). 

We now calculate the sum $A_{abc}h_{ab}$, where $h_{ab}\equiv
h_{ab}^{(1)}-h_{ab}^{(2)}$. Using the properties of coefficients 
$A_{abc}$ and solutions for $h_{ab}^{(1,2)}$ obtained above, one can 
find that 
\begin{eqnarray}
\label{differ2}
   A_{abc}h_{ab}&=&-\frac{1}{3}\sqrt{\frac{2}{3}}
                 (h_{88}-2h_{00}+3h_{11}+4h_{44})\delta_{c0}
                 \nonumber \\
                &-&\frac{2}{3\sqrt{3}}
                 (h_{88}+\sqrt{2}h_{08}-3h_{11}+2h_{44})\delta_{c8}
                   \nonumber \\
                &\equiv&H_1\delta_{c0}+H_2\delta_{c8}.
\end{eqnarray}
Again, from this equation the related formula (\ref{differ}) can be
established by equating $\omega_u=\omega_s$. In this special case we 
have $H_2=0$. As a next step let us contract the result with functions
$h_{ci}^{(1)}$
\begin{eqnarray}
   &&A_{abc}h_{ab}h_{ci}^{(1)}=\frac{1}{\sqrt{3}}
            \\
   &&       \left\{
          \begin{array}{ll}
          H_1(\sqrt{2}h^{(1)}_{00}+h^{(1)}_{08})+ 
          H_2(\sqrt{2}h^{(1)}_{08}+h^{(1)}_{88}), & i=u,d \\
    \\    H_1(\sqrt{2}h^{(1)}_{00}-2h^{(1)}_{08})+
          H_2(\sqrt{2}h^{(1)}_{08}-2h^{(1)}_{88}), & i=s
          \end{array}\right.
          \nonumber 
\end{eqnarray}
These contributions can be evaluated explicitly. It leads to the 
final expressions for the semiclassical corrections $z_i$ to the 
gap equation (\ref{gap}). They are given by
\begin{equation}
\label{gap-h}
          \begin{array}{l} 
          \displaystyle
          z_u=z_d \\
          \\
          \displaystyle
          =-\frac{\kappa\Omega^{-1}\omega_u}{8G^2\mu}
          \left[ 
          \frac{2(1-2\omega_u^2)-\omega_s}{(1-2\omega^2_u)^2-\omega_s^2}
          +\frac{2}{1-\omega_u^2}-\frac{3\omega_s}{1-\omega_s^2}
          \right], \\
              \\
          \displaystyle
          z_s=-\frac{\kappa\Omega^{-1}}{8G^2\mu}
          \left[ 
          \frac{4\omega^2_u(3\omega^2_u-2)+1+\omega_s}{(1-2\omega^2_u)^2
          -\omega_s^2}+\frac{3}{1-\omega_s} \right.\\
          \\
          \displaystyle\left. \ \ \ \ \ \ \ \ \ \ \ \ \ \ \ \ \ \ \ \
          -\frac{4}{1-\omega^2_u}
          \right],
          \end{array}
\end{equation}
where $\mu =(1+\omega_s-2\omega_u^2)$. Each of these formulas has 
the same limiting value at $\omega_s=\omega_u=\omega$ which coincides 
with Eq.(\ref{dif}). 

We apply this result to establish the $\hbar$-order correction to the
masses of constituent quarks. To this end we must use the general
expressions obtained in Appendices \ref{A1} and \ref{A2}, which 
for the considered case we rewrite in a way that stresses the 
quark content of the contributions
\begin{equation}
\label{Dmi}
   \Delta m_i=-G\sum_{j,k=1}^2z_j{\cal M}_{jk}{\cal Q}_{(i)k}
\end{equation}
where $i=u,s$, the $\hbar$-corrections $z_j$ are written as a line 
matrix $z_j=(z_u,z_s)$, and obviously $\Delta m_u=\Delta m_d$. 
The $2\times 2$ matrix ${\cal M}_{jk}$ and the column ${\cal Q}_{(i)k}$ 
are defined as follows
\begin{equation} 
  {\cal M}_{jk}=\left(
          \begin{array}{cc}
          1+\omega_s  &2\omega \\
          \omega      &1
          \end{array}
          \right), \qquad
   {\cal Q}_{(i)k}=\left(
          \begin{array}{c}
          {\cal Q}_{iu}+{\cal Q}_{id} \\
          {\cal Q}_{is}
          \end{array}
          \right)_k.
\end{equation}
Observe that $\mbox{det}{\cal M}=\mu$. The $\hbar$-corrections to the
quark masses must be calculated at the point $(M_u,\ M_s)$, being a 
solution of the gap equation at leading order. The expression 
(\ref{corr}) is a straightforward consequence of the more general 
result (\ref{Dmi}). Let us also note that formula (\ref{Dmi}) clearly 
shows which part of the correction is determined by the strange
component of the quark sea and which one by the non-strange 
contributions.

%%%%%%%%%%%%%%%%%%%%%%%%%%%%%%%%%%%%%%%%%%%%%%%%%%%%%%%%%%%%%%%%%%%%%%%%%%% 

\subsection{Effective potential: the $SU(2)_I\times U(1)_Y$ symmetry}
\label{ssF}
%%%%%%%%%%%%%%%%%%%%%%%%%%%%%%%%%%%%%%%%%%%%%%%%%%%%%%%%%%%%%%%%%%%%%%%%%%%
We can now generalize the result obtained in the Sec. \ref{effpot1}
to the case of $SU(2)_I\times U(1)_Y$ symmetry. To find the effective 
potential for this case one has to evaluate the line integral of the 
form
\begin{equation}
\label{int}
    \int_\gamma 2f_udm_u+f_sdm_s=V(\Delta_u, \Delta_s)
\end{equation}
where the independent variables $m_u$ and $m_s$ are linear combinations of 
the $SU(3)$ singlet and octet components $m_u=(\sqrt{2}m_0+m_8)/\sqrt{3}, 
\quad m_s=(\sqrt{2}m_0-2m_8)/\sqrt{3}$. We also know that   
\begin{eqnarray}
   f_u(m_u,m_s)&=&-\frac{h_u}{2}-\frac{N_c}{4\pi^2}m_uJ_0(m_u^2)
                 +\frac{z_u}{2}\ , 
                 \nonumber \\
   f_s(m_u,m_s)&=&-\frac{h_s}{2}-\frac{N_c}{4\pi^2}m_sJ_0(m_s^2)
                +\frac{z_s}{2}\ . 
\end{eqnarray}
The functions $f_u, f_s$ lie on the surface $S(m_u,m_s)$ defined by 
the stationary point equation (\ref{huhs}), and $\gamma$ is contained 
in $S$. There are some troubles caused by the singularities in $z_u, 
z_s$. The poles are located on curves which divide the surface 
$S(m_u,m_s)$ on distinct parts $\Sigma_n$. The integral
(\ref{int}) is well defined inside each of these regions $\Sigma_n$. 
It is characterized 
by the property that the integral over an arc $\gamma$, which is 
contained in $\Sigma_n$, depends only on its end points, i.e., the 
integral over any closed curve $\gamma_c$ contained in $\Sigma_n$ is 
zero. This follows from the fact that the integrand is an exact 
differential. Indeed, one can simply check that the one-form 
$2f_udm_u+f_sdm_s$ is closed on $\Sigma_n$:
\begin{equation}
   2\frac{\partial f_u}{\partial m_s}-\frac{\partial f_s}{\partial
          m_u}=0.
\end{equation}
On the other hand, the open set $\Sigma_n$ is diffeomorphic to ${\bf R}^2$ 
and, by Poincar\'e's lemma, the one-form is exact.    

Direct verification is relatively cumbersome and can be done along the 
lines of our calculation in the Appendix \ref{A3}, as follows. Since 
each of the differentials $2h_udm_u+h_sdm_s$ and $2z_udm_u+z_sdm_s$ is 
closed, let us consider them separately. We begin by evaluating the 
first one-form
\begin{equation}  
  -(2h_udm_u+h_sdm_s)
  =d\left(Gh_u^2+\frac{G}{2}h_s^2
  +\frac{\kappa}{8}h_u^2h_s\right)
\end{equation}
where we have used Eqs.(\ref{saddle-1}) to extract $dm_u$ and $dm_s$. 

Noting that $3\kappa^2(h_u^2+Q)=-\mu(16G)^2$, one can obtain for the
second one-form
\begin{equation}
\label{inter}
  2z_udm_u+z_sdm_s
  =\frac{3\kappa}{16}\left(z_s+2\omega_uz_u\right)dQ
  -2G\mu z_udh_u.
\end{equation}
Let us note also that
\begin{equation}
  z_s+2\omega_uz_u=-\frac{\kappa\Omega^{-1}\omega_s}{8G^2}\left[
  \frac{1}{(1-2\omega^2_u)^2-\omega_s^2}+\frac{3}{1-\omega_s^2}\right].
\end{equation}
Putting this expression in Eq.(\ref{inter}), we have after some algebra 
\begin{eqnarray}
  && 2z_udm_u+z_sdm_s \\
  \nonumber \\
  &=&\Omega^{-1}\left[\frac{\displaystyle 
   d\omega_s^2+4(1-2\omega_u^2)d\omega_u^2}{(1-2\omega_u^2)^2-\omega^2_s}
   +\frac{3d\omega^2_s}{1-\omega_s^2}
   +\frac{4d\omega_u^2}{1-\omega_u^2}\right]
   \nonumber \\
   \nonumber \\
  &=&-\Omega^{-1}d\ln\left|[(1-2\omega_u^2)^2-\omega_s^2]
   (1-\omega_s^2)^3(1-\omega^2_u)^4\right|.
       \nonumber 
\end{eqnarray}
Finally, we obtain the effective potential
\begin{equation}
\label{efpot2}
    \begin{array}{l}
    \displaystyle 
    V(\Delta_u,\Delta_s)=\frac{1}{2}\left(Gh_u^2+\frac{G}{2}h_s^2
             +\frac{\kappa}{8}h_u^2h_s\right)\\
    \displaystyle \ \ \ \ \ \ \ \ \ \ \ \ \ \ \ 
     -v(m_u^2)-\frac{1}{2}v(m_s^2)-C   \\
    \displaystyle   
    -\frac{\Omega^{-1}}{2}\ln\left|[(1-2\omega_u^2)^2-\omega_s^2]
              (1-\omega_s^2)^3(1-\omega^2_u)^4\right|
    \end{array}
\end{equation}
where $v(m_i^2)$ has been introduced in Eq.(\ref{v}).
The constant $C$ depends on the initial point of the curve $\gamma$,
and in the region $\Sigma$ which includes the point $\Delta_u=0, 
\Delta_s=0$ it can be fixed by requiring $V(0,0)=0$. This result 
coincides with Eq.(\ref{effpot}) in the $SU(3)$ limiting case.

\begin{figure}[t]
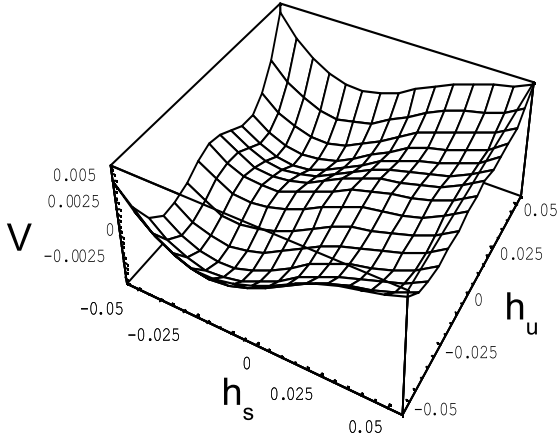

\PSfig{0.9}{fig.6}
\caption{Classical effective potential $V[\mbox{GeV}]$, 
Eq.(\ref{efpot2}) 
with $\Omega^{-1}=0$, as function of condensates $[\mbox{GeV}^{-3}]$
$h_u=2<\bar uu>$, $h_s=2<\bar ss>$ for $\Lambda =860$\ MeV,
$G=14\ \mbox{GeV}^{-2},\ \kappa =-1000\ \mbox{GeV}^{-5},\
\hat m_u=6\ \mbox{MeV}, \hat m_s=150\ \mbox{MeV}$.}
\label{fig6}
\end{figure}

In Fig.\ref{fig6} we show the effective potential calculated as 
function of condensates without the fluctuations, $\Omega^{-1}=0$,
for the parameter set given in the caption. There are altogether
nine critical points: four minima, four saddles and one maximum.
Only one critical point is localized in the region of physical 
interest, the minimum for $m_u, m_s>0$ (or $h_u, h_s<0$). In the 
chiral limit and otherwise the same parameters one has three 
critical points of interest, the maximum at the origin, the saddle 
at $m_u=0, m_s>0$ ($h_u=0, h_s<0$), and the minimum. This distribution
and behavior of critical points is common to a large set of parameters.

The behavior of $V(\Delta_u,\Delta_s)$ in terms of the strength of 
the fluctuation term $\sim\Omega^{-1}$ is qualitatively the same as 
for the $SU(3)$ case: fluctuations tend to restore the trivial vacuum 
in the region prior to the singularities. We also see from 
Eq.(\ref{efpot2}) that now the picture of singularities is more 
elaborated. Nevertheless attractive wells still develop between them.   

These results are in agreement with the following topological 
consideration. The effective potential $V$ is a smooth function
defined on the space of paths $S(h_u,h_s)$ diffeomorphic to 
${\bf R}^2$ (before the onset of caustics). The Euler characteristic 
of the surface $S$, $\chi (S)=1$, can be expressed, by Morse's 
theorem, through the number of non-degenerate critical points of the 
function $V$ 
\begin{equation}
   \chi (S)=C_0-C_1+C_2
\end{equation}
where $C_0$ is a number of critical points with index $0$ (minima),
$C_1$ is a number of critical points with index $1$ (saddle-points),
and $C_2$ is a number of critical points with index $2$ (maxima).

%%%%%%%%%%%%%%%%   FIG.6   %%%%%%%%%%%%%%%%%%%%%%%%%%%%%%%%%%%%%%%%%%%%%%%%%
%%%%%%%%%%%%%%%%%%%%%%%%%%%%%%%%%%%%%%%%%%%%%%%%%%%%%%%%%%%%%%%%%%%%%%%%%%%%%

%%%%%%%%%%%%%%%%%%%%%%%%%%%%%%%%%%%%%%%%%%%%%%%%%%%%%%%%%%%%%%%%%%%%%%%%%%%%%%%
\section{The ground state in the $1/N_c$ expansion}
\label{sec-4}
%%%%%%%%%%%%%%%%%%%%%%%%%%%%%%%%%%%%%%%%%%%%%%%%%%%%%%%%%%%%%%%%%%%%%%%%%%%%%%%
We have considered till now the semiclassical approach to estimate
the integral ${\cal Z}[\sigma ,\phi ;\Delta ]$ in Eq.(\ref{intJ}). 
However, rather than using $\hbar$ as the parameter of asymptotic 
expansion, we could also have used $1/N_c$ to estimate it.
In this case the stationary phase equations (\ref{saddle}) involve 
terms with different orders of $1/N_c$ and must be solved 
perturbatively. For this purpose let us represent (\ref{saddle}) 
in a complex form
\begin{equation}
\label{saddle-2}
  GU_a+W_a+\frac{3\kappa}{32}A_{abc}U^\dag_bU^\dag_c=0.
\end{equation}
Indeed the first two terms are of order $(1/N_c)^0$, since $G\sim 
1/N_c$, $U\sim N_c$, and $W\sim N_c^0$, while the last one is of order 
$1/N_c$. Casting the solutions $U_{\mbox{st}}^a$ as a series in
$1/N_c$ up to and including the terms of order $1/N_c$, we have 
\begin{equation}
\label{N_c-res}
   U^a_{\mbox{st}}=-\frac{1}{G}\left(W_a
                   +\frac{3\kappa}{32G^2}A_{abc}W^\dag_bW^\dag_c
                   +{\cal O}(1/N_c^2)\right). 
\end{equation}
It yields for $ {\cal L}_r(r_{\mbox{st}})$
\begin{eqnarray}
\label{lam-Nc}
   {\cal L}_r(r_{\mbox{st}})&=&-\frac{1}{4G}\mbox{tr}(WW^\dag )
            -\frac{\kappa}{(4G)^3}\left(\det W+\det W^\dag\right)
            \nonumber \\   
            &+&{\cal O}(1/N_c). 
\end{eqnarray}  

The contribution from the auxiliary fields can be obtained directly 
from Eq.(\ref{Sra}) by expanding our solutions $h_{ab}$ and $h_c$ in a 
series in $1/N_c$. One can already conclude from that expression, 
without any calculations, that the term $\sim\Omega^{-1}$ is at most of 
order $\sim 1/N_c$. Therefore it is beyond the accuracy of the considered
approximation and can be neglected. Actually, it follows from 
Eq.(\ref{ha}), that the difference 
$h_{ab}=h_{ab}^{(1)}-h_{ab}^{(2)}$ has order $\sim N_c^0$, additionally
suppressing this contribution, i.e. for the correction to the result  
(\ref{lam-Nc}) we have only a term starting from $1/N_c^2$-order
\begin{eqnarray}
\label{lam-Nc+}
   \Delta{\cal L}_r&=&-\frac{\kappa^2\Omega^{-1}}{8(2G)^4}
         \left[\mbox{tr}(WW^\dag )
         +\frac{3\kappa}{(4G)^2}
         \left(\det W+\det W^\dag\right)\right]
         \nonumber \\
         &+&{\cal O}(1/N_c^3). 
\end{eqnarray}  
It means, in particular, that one can neglect this type of quantum
fluctuations in the discussion of the gap equation up to and including
the terms of $1/N_c$ order in the meson Lagrangian. These corrections
cannot influence significantly the dynamical symmetry breaking
phenomena in the model and only the quantum effect of mesons (the 
one-loop contributions of $\sigma$ and $\phi$ fields) together with
the leading order contribution from the 't Hooft determinant 
(see Eq.(\ref{lam-Nc})) are relevant at $N_c^0$-order here.

If the model allows us to utilize the $1/N_c$ expansion, the vacuum 
state is defined by the gap equation obtained from Eq.(\ref{gap}) in
the large $N_c$ limit, or, equivalently, on the basis of
Lagrangian (\ref{lam-Nc}). We have
\begin{equation}
\label{gap-Nc}
  \Delta_i
  +\frac{\kappa}{32G^2}t_{ijk}\Delta_j\Delta_k
  =\frac{N_cG}{2\pi^2}m_iJ_0(m_i).
\end{equation}
At leading order in an $1/N_c$ expansion we have the standard 
gap equation $2\pi^2\Delta_i =N_cGm_iJ_0(m_i)$. The terms arising 
on the next step already include the quantum correction from the 
't Hooft determinant. It is not difficult to obtain the corresponding
contribution of order $1/N_c$ to the mass of constituent quarks
\begin{eqnarray}
   &&m_i=M_i+\Delta m_i\ , 
         \nonumber \\
         \nonumber \\
   &&\Delta m_i=-\frac{\kappa \Delta_j\Delta_k}{
       (4G)^2(1-I_i)} \qquad (i\neq j\neq k)
\end{eqnarray}
where $\Delta m_i$ is calculated at the point $m_i=M_i$, which is the 
solution of the gap equation (\ref{gap-Nc}) at leading order, 
and $I_i$ is given by Eq.(\ref{I_i}). One can show that $\Delta m_i>0$, 
thus increasing the effect of the dynamical chiral symmetry breaking. 
This is an immediate consequence of the formula
\begin{equation}
   1-I_i(M_i^2)
   =\frac{\hat{m}_i}{M_i}+\frac{N_cG}{\pi^2}M_i^2J_1(M_i^2)>0.
\end{equation}
Let us stress that the first equality is fulfilled only at the point 
$M_i$.

We did not clarify yet the counting rule for the current quark masses
$\hat{m}_i$, assuming that they are counted as the constituent quark 
masses.
Actually, these masses are small and following the standard rules of
ChPT one should consider $\hat{m}_i\sim 1/N_c$. In this case in the large
$N_c$ limit the model possesses the $U(3)\times U(3)$ symmetry and the
gap equation at leading order, $2\pi^2=N_cGJ_0(m_i)$, leads to a
solution with equal masses $M_u=M_d=M_s\equiv M$. The $1/N_c$ 
correction includes the 
$\kappa$-dependent term and the term depending on the current quark masses
\begin{equation}
   \Delta m_i=\frac{\pi^2}{N_cG}\left(\hat{m}_i-\frac{\kappa M^2}{16G^2}
              \right)\frac{1}{M^2J_1(M^2)}\ .
\end{equation}

Returning back to Eq.(\ref{lam-Nc}) one can conclude that the large-$N_c$
limit corresponds to the picture which is not affected by six-quark
fluctuations. This can be realized also directly from Lagrangian
(\ref{lam}). Indeed, the couplings of meson vertices are
determined here through the functions $h_a$ given in the case of
$SU(3)$ flavor 
symmetry by $h_u^{(1)}$ in Eq.(\ref{u=d=s}). The large-$N_c$ limit
forces a series expansion for $h_u^{(1)}$ with a small parameter
\begin{equation}
   \epsilon =\frac{|\kappa |\Delta_u}{4G^2}\sim\frac{1}{N_c}\ll 1 
\end{equation} 
and the leading term $h_u^{(1)}=-\Delta_u/G+\ldots$ which does not
depend on $\kappa$. 

This observation leads us to the second important conclusion. It is 
easy to see that the parameter $\epsilon$ is an internal model
parameter and the series expansion in $\epsilon$ closely corresponds 
to the $1/N_c$ expansion of the model. The existence of this small 
parameter allows us to consider the $1/N_c$ series as a perfect 
approximation for the system with {\it small} vacuum six-quark 
fluctuations
\begin{equation}
   |\kappa|\ll\frac{4G^2}{\Delta_u}\ ,
   \qquad \Delta_u=\Delta_u(G,\Lambda ). 
\end{equation}  

What to do if $\epsilon$ is not too small? A large value for 
$\epsilon$ can simply distabilize the $1/N_c$ series, implying 
large $1/N_c$ corrections. It has been observed recently 
\cite{Pennington:1998} that the abundance of strange
quark-antiquark pairs in the vacuum can lead to nonnegligible vacuum
correlations between strange and non-strange quark pairs. If this 
happens one can try to understand the behaviour of the quark
system on the basis of the $\hbar$-expansion. This approximation can
be considered as the limit of {\it large} six-quark fluctuations. 
Although QCD does not contain an obvious parameter which
could allow one to describe this limit, the model under consideration,
as one can see, for instance, from Eq.(\ref{trivvac}),
suggests this dimensionless parameter:
\begin{equation}
   \zeta =\frac{\kappa^2\Omega^{-1}}{32G^3}\ll 1.
\end{equation}  
On the basis of this inequality one can conclude that values of $\kappa$, 
corresponding to large six-quark fluctuations are determined
by the condition
\begin{equation}
  |\kappa |\ll 2\sqrt{(2G)^3\Omega }.
\end{equation}
One can see that $|\kappa |_l\sim\sqrt{N_c}|\kappa |_s$, where we used
letters $l,s$ to mark possible values of $|\kappa |$ for large and
small six-quark fluctuations correspondingly. These two regimes lead 
to the different patterns of chiral symmetry breaking. In the first case 
the mass of $\eta'$ meson goes to zero when $N_c\rightarrow\infty$. 
In the second case the ground state does not have the $\eta'$ Goldstone
boson even at leading order.

%%%%%%%%%%%%%%%%%%%%%%%%%%%%%%%%%%%%%%%%%%%%%%%%%%%%%%%%%%%%%%%%%%%%%%%%%%%%%%%
\section{Concluding remarks}
\label{sec-5}
%%%%%%%%%%%%%%%%%%%%%%%%%%%%%%%%%%%%%%%%%%%%%%%%%%%%%%%%%%%%%%%%%%%%%%%%%%%%%%%
The purpose of this paper has been to use the path integral approach
to study the vacuum state and collective exitations of the 't Hooft
six quark interaction. We started from the bosonization procedure
following the technique described in the papers 
\cite{Diakonov:1986,Reinhardt:1988}. The leading order stationary phase 
approximation made in the path integral leads to the same result as
obtained by different methods, based either on the Hartree -- Fock 
approximation \cite{Bernard:1988}, or on the standard mean field 
approach \cite{Hatsuda:1994}. The stationary trajectories are
solutions for the system of the stationary phase equations and we 
find them in analytical form for the two cases corresponding to $SU(3)$
and $SU(2)_I\times U(1)_Y$ flavor symmetries. The exact knowledge
of the stationary path is a nesessary step to obtain the effective
bosonic Lagrangian. We give a detailed analytical solution to this 
problem. 

As the next step in evaluating the functional integral we have considered 
the semiclassical corrections which stem from the Gaussian integration. 
We have found and analyzed the corresponding contributions to the 
effective potential $V$, masses of constituent quarks and quark 
condencates. The most interesting conclusions are the following:
 
   (a) We have found that already the classical effective potential 
       $V(\Delta_u)$ for the case in which chiral symmetry 
       $SU_L(3)\times SU_R(3)$ is broken down to the $SU(3)$ subgroup, 
       has a metastable vacuum state, although the values
       of parameters $G, \kappa , \hat{m}_i, \Lambda$ corresponding
       to this pattern are quite unnatural from the physical point
       of view: the couplings $G$ and $\Lambda$ must be small  
       to fulfill the inequality $N_c\Lambda^2G\leq 2\pi^2$, which
       is known in the NJL model without the 't Hooft interaction 
       as a condition for the trivial vacuum; the coupling $\kappa$ 
       must be several times bigger (in absolute value) of the value 
       known from the instanton picture. Besides, the window of 
       parameters for the existence of metastable vacua
       is quite small. We show then that 
       semiclassical corrections, starting from some increasing
       critical value of the strength $\Lambda_E$,  
       transform any classical potential with a single spontaneously 
       broken vacuum to the semiclassical potential with a single
       trivial vacuum. Close to the chiral limit this transition goes 
       through a smooth sequence of potentials with two minima. 
There are other known cases of effective chiral Lagrangians which 
confirm the picture with several vacua \cite{Veneziano:1980,Halperin:1998}.    
   
   (b) If the symmetry is broken up to the $SU(2)_I\times U(1)_Y$ 
       subgroup the smooth classical effective potential 
       $V(\Delta_u,\Delta_s)$ defined on the space of stationary 
       trajectories $S$ may have several non-degenerate critical 
       points. It is known that the properties of the critical 
       points are related to the topology of the surface $S$. We
       used this geometrical aspect of the problem to draw conclusions 
       about an eventual more elaborate structure of the hadronic 
       vacuum already at leading order in $\hbar$. We find for some 
       parameter sets the existence of a minimum, a maximum and a 
       saddle point. For vanishing current quark masses, for example, 
       the minimum corresponds to the spontaneous breakdown of chiral 
       symmetry, the maximum is at the origin, and the saddle at 
       $m_u\sim 0$ and $m_s$ finite. Similar as in the $SU(3)$
       case, the inclusion of fluctuations tends to destroy the
       spontaneous broken phase and to restore the trivial vacuum.

Our work raises some issues which can be addressed and used in further 
calculations:

(1) The Gaussian approximation leads to singularities in the
    effective potential. To study this problem which is known as 
    caustics in the path integral, it is necessary to go beyond 
    this approximation and take into account quantum fluctuations
    of higher order than the quadratic ones. 
    To the level of accuracy of the WKB approximation the effective
    potential is elsewhere well-defined and has stable minima between
    the singularities. It is interesting to trace the fate of the
    these minima going beyond the WKB approximation, since 
    in this work we have analyzed the 
    effective potential mainly at a safe distance from
    the first caustic. 

(2) Our expressions for the mass corrections (\ref{deltami})
    have a general form and
    can be used, for instance, to include the one-loop effect of 
    mesonic fields. One can use it as well to find the relative 
    strength of strange and non-strange quark pairs in this 
    contribution.

(3) We have chosen $\hbar$ and independently $1/N_c$ as two possible 
    parameters for the systematic expansion of the effective action.
    As we saw the $1/N_c$ expansion is a much more restrictive
    procedure. They are of interest for the study of the lowest lying
    scalar and pseudoscalar meson spectrum. There is a qualitative
    understanding of this spectrum at phenomenological level 
    (see, for instance, papers \cite{Hooft:1999}).
    A more elaborate study might lead to the necessity of including
    either additional many-quark vertices or taking into account 
    systematically quantum corrections. Our results might be
    helpful in approaching both of the indicated developments.

%%%%%%%%%%%%%%%%%%%%%%%%%%%%%%%%%%%%%%%%%%%%%%%%%%%%%%%%%%%%%%%%%%%%%%%%%%%%%%%
\begin{acknowledgement}
We are very grateful for discussions with A. Pich, G. Ripka and N. Scoccola
during the II International Workshop on Hadron Physics, Coimbra, 
Portugal, where
part of the material of this work has been presented.
This work is supported by grants provided by Funda\ca o para a 
Ci\^encia e a Tecnologia, POCTI/35304 /FIS/2000.
\end{acknowledgement}
%%%%%%%%%%%%%%%%%%%%%%%%%%%%%%%%%%%%%%%%%%%%%%%%%%%%%%%%%%%%%%%%%%%%%%%%%%%%%%%

\appendix
%%%%%%%%%%%%%%%%%%%%%%%%%%%%%%%%%%%%%%%%%%%%%%%%%%%%%%%%%%%%%%%%%%%%%%%%%%%%%%
\section{The semiclassical corrections to the constituent quark masses}
\label{A1}
%%%%%%%%%%%%%%%%%%%%%%%%%%%%%%%%%%%%%%%%%%%%%%%%%%%%%%%%%%%%%%%%%%%%%%%%%%%%%
To derive the explicit formula for the semiclassical next to the
leading order corrections $\Delta m_i$ to the constituent quark 
masses $M_i$ one has to solve the system of equations, following 
from Eq.(\ref{gap}) 
\begin{equation}
\label{delm-i}
          \left\{
          \begin{array}{r} 
          \displaystyle
          \frac{I_u}{G}\Delta m_u+\sum_{i=u,d,s}
          \frac{\partial h_u}{\partial\Delta_i}\Delta m_i=z_u
          \\
          \displaystyle
          \frac{I_d}{G}\Delta m_d+\sum_{i=u,d,s}
          \frac{\partial h_d}{\partial\Delta_i}\Delta m_i=z_d
          \\
          \displaystyle
          \frac{I_s}{G}\Delta m_s+\sum_{i=u,d,s}
          \frac{\partial h_s}{\partial\Delta_i}\Delta m_i=z_s
          \end{array}\right.
\end{equation}
where the functions $I_i(m_i^2)$ are given by Eq.(\ref{I_i}). Both the
partial derivatives and integrals $I_i(m_i^2)$ must be calculated for 
$m_i=M_i$. From the first system of equations in (\ref{order-h0}) we may 
express derivatives $\partial h_i/\partial\Delta_j$ in terms of the 
functions $h_i$. To symplify the work it is convenient to change the
notation and rewrite (\ref{saddle-1}) in the form 
\begin{equation}
\label{sadd}
          \left\{
          \begin{array}{r} 
          \displaystyle
          \omega_u+x_u+\omega_d\omega_s=0
          \\
          \displaystyle
          \omega_d+x_d+\omega_u\omega_s=0
          \\
          \displaystyle
          \omega_s+x_s+\omega_u\omega_d=0
          \end{array}\right. \quad
          \omega_i\equiv\frac{\kappa h_i}{16G}\ ,\quad
          x_i\equiv\frac{\kappa\Delta_i}{(4G)^2}\ .
\end{equation}
Straightforward algebra on the basis of these equations gives
\begin{equation}
   \frac{\partial\omega_i}{\partial x_i}=-\frac{1}{A}(1-\omega_i^2),
   \qquad
   \frac{\partial\omega_i}{\partial x_j}=-\frac{1}{A}
   (\omega_i\omega_j-\omega_k)=
   \frac{\partial\omega_j}{\partial x_i}\ .
\end{equation}
Here $A=1-\omega_u^2-\omega_d^2-\omega_s^2+2\omega_u\omega_d\omega_s$.
We assume that indices $i,j,k$ range over the set $\{u,d,s\}$ in
such a way that $i\neq j\neq k$, and a sum over repeated index is
implied only if the symbol of the sum is explicitly written. 

The main determinant of the system (\ref{delm-i}) is equal to
\begin{eqnarray}
   {\cal D}&=&-\frac{D}{AG^3}\ ,\qquad D=1-\sum_{i=u,d,s}I_i+
   (1-\omega_u^2)I_dI_s
   \nonumber \\
   &+&(1-\omega_d^2)I_sI_u+(1-\omega_s^2)I_uI_d
   -AI_uI_dI_s
\end{eqnarray}
with $I_i$ given by Eq.(\ref{I_i}).
The other related determinants are written in the compact form 
\begin{equation}
\label{D_i}
   {\cal D}_i=\frac{1}{AG^2}\left(
   z_iB_{jk}+z_jB_{jik}+z_kB_{kij}\right), \qquad (i\neq j\neq k)
\end{equation}
where 
\begin{eqnarray}
  && B_{ij}=1-(1-\omega_i^2)I_j-(1-\omega_j^2)I_i+AI_iI_j, 
           \nonumber \\
  && B_{ijk}=(\omega_i\omega_j-\omega_k)I_k+\omega_k.
\end{eqnarray}
Hence the $\hbar$-correction to the mean field value of the
constituent quark mass is given by
\begin{equation}
\label{deltami}
   \Delta m_i=\frac{{\cal D}_i}{{\cal D}}=-\frac{G}{D}
   \left(z_iB_{jk}+z_jB_{jik}+z_kB_{kij}\right) 
\end{equation}
where $i\neq j\neq k$.
This formula gives us the most general expression which has to be
specified by the explicit form for $z_i$. 

Let us also write out the two partial cases for this result. If 
$\omega_u=\omega_d\equiv\omega$ and $I_u=I_d=I$, which happens when 
the group of flavor symmetry is broken according to the pattern 
$SU(3)\rightarrow SU(2)_f\times U(1)_f$, one can find 
\begin{eqnarray} 
     A&=&(1-\omega_s)\mu , \qquad \mu\equiv (1+\omega_s-2\omega^2),
     \nonumber \\
     D&=&[1-(1-\omega_s)I][1-I_s-(1+\omega_s)I+\mu II_s],  
     \nonumber \\ 
     AG^2{\cal D}_u&=&[1-(1-\omega_s)I][z_u(1+\omega_s-\mu
     I_s)+z_s\omega ], 
     \nonumber \\
     AG^2{\cal D}_s&=&[1-(1-\omega_s)I][z_s(1-\mu I)+2z_u\omega ]. 
\end{eqnarray}
This determines the coefficients in $\Delta m_i$, and we obtain the
following form of mass corrections
\begin{eqnarray}
\label{dmsu2}
   \Delta m_u&=&G\frac{z_u\mu I_s-z_u(1+\omega_s)-z_s\omega}{
                1-I_s-(1+\omega_s)I+\mu II_s}\ , 
                \nonumber \\
   \Delta m_s&=&G\frac{z_s(\mu I-1)-2z_u\omega}{
                1-I_s-(1+\omega_s)I+\mu II_s}\ .
\end{eqnarray}
The second partial case for the formula (\ref{deltami}) corresponds to
the $SU(3)$ flavor symmetry. It is clear that now we have 
\begin{eqnarray} 
     && A=(1+2\omega )(1-\omega )^2,
        \nonumber \\
     && D=[1-(1+2\omega )I][1-(1-\omega )I]^2, 
        \nonumber \\
     && AG^2{\cal D}_u=z_u(1+2\omega )[1-(1-\omega )I]^2.
\end{eqnarray}
Then (\ref{deltami}) yields the following result for the mass correction 
\begin{equation}
\label{dmsu3}
   \Delta m_u=-G\frac{z_u(1+2\omega )}{1-(1+2\omega )I}\ .
\end{equation}

%%%%%%%%%%%%%%%%%%%%%%%%%%%%%%%%%%%%%%%%%%%%%%%%%%%%%%%%%%%%%%%%%%%%%%%%%%%%%%
\section{Particle expectation values ${\cal Q}_{ij}$}
\label{A2}
%%%%%%%%%%%%%%%%%%%%%%%%%%%%%%%%%%%%%%%%%%%%%%%%%%%%%%%%%%%%%%%%%%%%%%%%%%%%%
The knowledge of the constituent quark mass $M_i$ gained from the 
equations (\ref{order-h0}), combined together with the Feynman -- 
Hellmann theorem \cite{Feynman:1939}, have been used for finding 
the expectation values of the scalar quark densities, $\bar{q}_jq_j$, 
in the physical quark state $|Q_i>$ (see paper \cite{Bernard:1988}). 
The matrix element $<Q_i|\bar{q}_jq_j|Q_i>$ describes the mixing of 
quarks of flavor $j$ into the wavefunction of constituent quarks of 
flavor $i$. One can determine these particle expectation values by 
calculating the partial derivatives 
\begin{equation}
\label{pev}
   <Q_i|\bar{q}_jq_j|Q_i>=\frac{\partial M_i}{\partial\hat{m}_j}
   \equiv {\cal Q}_{ij}. 
\end{equation}
The functions ${\cal Q}_{ij}$ have been calculated in \cite{Bernard:1988} 
for the case $\hat{m}_u=\hat{m}_d$. 
There are a number of physical problems in which these matrix elements
are usefull. We have found the presence of them in the expression for 
the semiclassical corrections to the mean field quark masses $M_i$. 

Both the mass corrections $\Delta m_i$ and the
particle expectation values ${\cal Q}_{ij}$ are solutions of a
similar system of equations and can be derived on an equal footing.
Indeed, ${\cal Q}_{ij}$ are the solutions of the equations obtained from 
Eq.(\ref{order-h0}) by differentiation with respect to $\hat{m}_j$. 
These equations differ from the system 
(\ref{delm-i}) only up to the replacements of variables $\Delta m_i
\rightarrow{\cal Q}_{ij}$ and $z_i\rightarrow\partial h_i/\partial\Delta_j$.
Therefore we have
\begin{equation}
   {\cal Q}_{ij}=\frac{{\cal D}_{i(j)}}{{\cal D}}.
\end{equation}
With this way of writing the determinants ${\cal D}_{i(j)}$ we 
wish to stress that one can simply obtain them from ${\cal D}_i$ in 
Eq.(\ref{D_i}) through the above replacements $z_i\rightarrow\partial 
\omega_i/(G\partial x_j)$. The main determinant ${\cal D}$ is not 
changed. By use of these formulas we are led to the explicit
expressions  
\begin{eqnarray}
\label{pevg}
   {\cal Q}_{ii}&=&\frac{1}{D}\left[(1-I_j)(1-I_k)-\omega_i^2I_jI_k\right],
                   \nonumber \\
   {\cal Q}_{ij}&=&\frac{1}{D}I_iB_{ijk}, 
   \quad i\neq j\neq k \quad (\mbox{no sum})
\end{eqnarray}
which correspond to the most general case 
$\hat{m}_u\neq\hat{m}_d\neq\hat{m}_s$. 
The notation have been explained in Appendix \ref{A1}.

%%%%%%%%%%%%%%%%%%%%%%%%%%%%%%%%%%%%%%%%%%%%%%%%%%%%%%%%%%%%%%%%%%%%%%%%%%%%%%
\section{Effective potential}
\label{A3}
%%%%%%%%%%%%%%%%%%%%%%%%%%%%%%%%%%%%%%%%%%%%%%%%%%%%%%%%%%%%%%%%%%%%%%%%%%%%%
The models which are considered here lead in the most general case to 
the potential $V(\Delta_u,\Delta_d,\Delta_s)=V(\Delta_i)$. This
function is a solution of Eq.(\ref{der}). One can reconstruct 
$V(\Delta_i)$ by integrating the one-form
\begin{equation}
\label{dV}
   dV=\frac{\partial V}{\partial m_u}dm_u
     +\frac{\partial V}{\partial m_d}dm_d
     +\frac{\partial V}{\partial m_s}dm_s=
     \sum_{i=u,d,s}f_idm_i.
\end{equation}
Thus, the derivation of the effective potential in the $f_u=f_d=f_s$
case is simply a question of representing the gap equation 
$f(m_u)=0$ in the form of an exact differential $3f(m_u)dm_u=dV(m_u)$, 
where $m_u=\sqrt{2/3}m_0$. One should also take into account  
the constraint  
\begin{equation}
\label{sph}
   Gh_u+\Delta_u+\frac{\kappa}{16}h_u^2=0,
\end{equation}
which implies that
\begin{equation}
\label{difsph}
   dm_u=-\left(G+\frac{\kappa}{8}h_u\right)dh_u
       =-\frac{16G^2}{\kappa}(1+2\omega )d\omega .
\end{equation}
Using this result one can obtain for the first term in (\ref{f}), 
for instance,
\begin{equation}
\label{1term}
   -h_u dm_u=h_u\left(G+\frac{\kappa}{8}h_u\right)dh_u
            =d\left(\frac{G}{2}h_u^2+\frac{\kappa}{24}h_u^3\right).
\end{equation}
A similar calculation with the third term in (\ref{f}) gives
\begin{eqnarray}
\label{3term}
    && \frac{\kappa\Omega^{-1}}{2G^2}
       \frac{\omega (1-3\omega^2)dm_u}{(1+2\omega)
       (1-4\omega^2)(1-\omega^2)}
       \nonumber \\
   &=&-4\Omega^{-1}\frac{(1-3\omega^2)d\omega^2}{(1-4\omega^2)(1-\omega^2)}
       \nonumber \\
   &=&\frac{1}{3}\Omega^{-1}d\ln\left|(1-4\omega^2)(1-\omega^2)^8\right|.
\end{eqnarray}
To conclude the procedure we must then add to the effective potential 
$V(\Delta_u)$ the corresponding contribution from the second term
\begin{equation}
\label{psi}
   -\frac{N_c}{2\pi^2}m_uJ_0(m_u^2)dm_u=-dv(m_u^2)
\end{equation}
where we have defined
\begin{equation}
\label{v}
   v(m_i^2)\equiv\frac{N_c}{8\pi^2}
   \left[m_i^2J_0(m_i^2)+\Lambda^4\ln\left(1+\frac{m_i^2}{\Lambda^2}
   \right)\right].
\end{equation} 
All this amounts to calculate the effective potential in the form 
given by Eq.(\ref{effpot}).

%%%%%%%%%%%%%%%%%% REFERENCES %%%%%%%%%%%%%%%%%%%%%%%%%%%%%%%%%%%%%%%%%%%%%%%%%
\baselineskip 12pt plus 2pt minus 2pt


\begin{thebibliography}{99}
%%%%%%%%%%%%%%%%%%%%%%%%%%%%%%%%%%%%%%%%%%%%%%%%%%%%%%%%%%%%%%%%%%%%%%%%%%%%%%%
\bibitem{Hooft:1976} A. M. Polyakov, Phys. Lett. B {\bf 59} (1975) 82;
        Nucl. Phys. B {\bf 120} (1977) 429; 
        A. A. Belavin, A. M. Polyakov, A. Schwartz and Y. Tyupkin,
        Phys. Lett. B {\bf 59} (1975) 85;
        G. 't Hooft, Phys. Rev. Lett. {\bf 37} (1976) 8; Phys. Rev. D 
        {\bf 14} (1976) 3432;
        C. Callan, R. Dashen and D. J. Gross, Phys. Lett. B {\bf 63}
        (1976) 334;
        R. Jackiw and C. Rebbi, Phys. Rev. Lett. {\bf 37} (1976) 172;
        S. Coleman, {\it The uses of instantons} (Erice Lectures, 1977). 
\bibitem{Diakonov:1995} D. Diakonov, {\it Chiral symmetry breaking by 
        instantons} (Lectures at the Enrico Fermi School in
        Physics, Varenna, June 27 - July 7, 1995) [arXiv:hep-ph/9602375].
\bibitem{Nambu:1961} Y. Nambu and G. Jona-Lasinio, Phys. Rev. {\bf 122}
        (1961) 345; {\bf 124} (1961) 246;
        V. G. Vaks and A. I. Larkin, Zh. \'{E}ksp. Teor. Fiz. {\bf 40} 
        (1961) 282.
\bibitem{Bernard:1988} V. Bernard, R. L. Jaffe and U.-G. Mei\ss ner, 
        Nucl. Phys. B {\bf 308} (1988) 753.
\bibitem{Kunihiro:1988} T. Kunihiro and T. Hatsuda, Phys. Lett. B {\bf
        206} (1988) 385;
        T. Hatsuda, Phys. Lett. B {\bf 213} (1988) 361;
        Y. Kohyama, K. Kubodera and M. Takizawa, Phys. Lett. B 
        {\bf 208} (1988) 165; 
        M. Takizawa, Y. Kohyama and K. Kubodera, Prog. Theor. Phys. 
        {\bf 82} (1989) 481. 
\bibitem{Diakonov:1986} D. Diakonov and V. Petrov, {\it Spontaneous
        breaking of chiral symmetry in the instanton vacuum} (preprint
        LNPI-1153, 1986, published (in Russian) in Hadron matter
        under extreme conditions, Kiev, 1986, p. 192);
        D. Diakonov and V. Petrov, in {\it Quark cluster dynamics}
        (Lecture Notes in Physics, Springer-Verlag, Vol. {\bf 417},
        p. 288, 1992).
\bibitem{Diakonov:1998} D. Diakonov, {\it Chiral quark-soliton model} 
        (Lectures at the Advanced Summer School on non-perturbative
        field theory, Peniscola, Spain,
        June 2-6, 1997) [arXiv:hep-ph/9802298]. 
\bibitem{Reinhardt:1988} H. Reinhardt and R. Alkofer, Phys. Lett. B
        {\bf 207} (1988) 482.
\bibitem{Osipov:2002} A. A. Osipov and B. Hiller, Phys. Lett. B {\bf 539}
        (2002) 76 [arXiv:hep-ph/0204182].
\bibitem{Kleinert:2000} H. Kleinert and B. Van den Bossche,
        Phys. Lett. B {\bf 474} (2000) 336 [arXiv:hep-ph/9907274].
\bibitem{Babaev:2000} E. Babaev, Phys. Rev. D {\bf 62} (2000) 074020
        [arXiv:hep-ph/0006087]; 
        T. Lee and Y. Oh, Phys. Lett. B {\bf 475} (2000) 207 
        [arXiv:nucl-th/9909078];
        T. Kashiwa and T. Sakaguchi, Preprint KYUSHU-HET 67 (2003) 
        [arXiv:hep-th/0306008]. 
\bibitem{Schulman:1981} L. S. Schulman, {\it Techniques and applications
        of path integration} (A Wiley-Interscience Publication,
        John Wiley \& Sons, New York, 1981). 
\bibitem{Kashiwa:2003} T. Kashiwa and T. Sakaguchi, Preprint 
        KYUSHU-HET 64 (2003) [arXiv:hep-th/0301019].
\bibitem{Osipov:2001} A.A. Osipov and B. Hiller, Phys. Rev. D {\bf 62}
        (2000) 114013 [arXiv:hep-ph/0007102]; Phys. Rev. D {\bf 63} 
        (2001) 094009 [arXiv:hep-ph/0012294].
\bibitem{Bogoliubov:1980} N. N. Bogoliubov and D. V. Shirkov,
        {\it Introduction to the theory of quantized fields}
        (A Wiley-Interscience Publication, John Wiley \& Sons, New York, 
        1980); S. Weinberg, {\it The quantum theory of fields}
        (Cambridge University Press, Cambridge, 1995). 
\bibitem{Hatsuda:1994} T. Hatsuda and T. Kunihiro, Phys. Rep. {\bf 247}
        (1994) 221 [arXiv:hep-ph/9401310].
\bibitem{Coleman:1988} S. Coleman, {\it Aspects of symmetry} 
        (Selected Erice Lectures, Cambridge University Press,
        New York, 1985).
\bibitem{Pennington:1998} M. R. Pennington [arXiv:hep-ph/9811276];
        S. Spanier and N. Tornqvist in K. Hagiwara {\it et al}.
        [Particle Data Group Collaboration], Phys. Rev. D {\bf 66}
        (2002) 010001.
\bibitem{Veneziano:1980} P. Di Vecchia and G. Veneziano,
        Nucl. Phys. {\bf B171} (1980) 253; 
        K. Kawarabayashi and N. Ohta, Nucl. Phys. B {\bf 175} (1980) 477; 
        E. Witten, Ann. Phys. (N.Y.) {\bf 128} (1980) 363;
        C. Rosenzweig, J. Schechter and C. G. Trahern, Phys. Rev. D
        {\bf 21} (1980) 3388;
        R. Arnowitt and Pran Nath, Phys. Rev. D {\bf 23} (1981) 473;
        Nucl. Phys. B {\bf 209} (1982) 234, 251.
\bibitem{Halperin:1998} M. Creutz, Phys. Rev. D {\bf 52} (1995) 2951 
        [arXiv:hep-th/9505112]; 
        I. Halperin and A. Zhitnitsky, Phys. Rev. Lett. {\bf 81}  
        (1998) 4071 [arXiv: hep-ph/9803301]; 
        A. V. Smilga, Phys. Rev. D {\bf 59} (1999) 114021        
        [arXiv:hep-ph/9805214]; 
        M. Shifman, Phys. Rev. D {\bf 59} (1999) 021501
        [arXiv:hep-th/9809184]; 
        M. H. Tytgat, Phys. Rev. D {\bf 61} (2000) 114009
        [arXiv:hep-ph/9909532]; 
        G. Gabadadze and M. Shifman, Int. J. Mod. Phys. A {\bf 17} 
        (2002) 3689 [arXiv:hep-ph/0206123]; 
        P. J. A. Bicudo, J. E. F. T. Ribeiro and A. V. Nefediev,
        Phys. Rev. D {\bf 65} (2002) 085026 [arXiv:hep-ph/0201173].
\bibitem{Hooft:1999} G. 't Hooft, arXiv:hep-th/9903189;
        M. Napsuciale and S. Rodriguez, Int. J. Mod. Phys. A {\bf 16} 
        (2001) 3011 [arXiv:hep-ph/0204149].
\bibitem{Feynman:1939} R. P. Feynman, Phys. Rev. {\bf 56} (1939) 340. 
\end{thebibliography}
\end{document}